\documentclass[aps,superscriptaddress,preprint]{revtex4-1}

\usepackage[table]{xcolor}
\usepackage{array}
\usepackage{color}
\usepackage[]{graphicx}
\usepackage{latexsym}
\usepackage{natbib}
\usepackage{amsmath}
\usepackage[caption=false]{subfig}
\usepackage{mathtools}
\usepackage{textcomp}
\usepackage{rotating}
\usepackage{tikz}
\usepackage{listings}
\usepackage{amssymb}

\begin{document}

\title{Symmetry group factorization reveals the structure-function
  relation in the neural connectome of {\it Caenorhabditis elegans} }

\author{Flaviano Morone}

\affiliation{Levich Institute and Physics Department, City College of
  New York, New York, NY 10031}

\author{Hern\'an A. Makse}

\affiliation{Levich Institute and Physics Department, City College of
  New York, New York, NY 10031}

\begin{abstract}

{\bf The neural connectome of the nematode {\it Caenorhabditis elegans} 
has been completely mapped, yet in spite of being one of the smallest 
connectomes (302 neurons), the design principles that explain how the 
connectome structure determines its function remain unknown. Here, 
we find symmetries in the locomotion neural circuit of {\it C. elegans}, 
each characterized by its own symmetry group which can be factorized 
into the direct product of normal subgroups. The action of these normal 
subgroups partitions the connectome into sectors of neurons that match 
broad functional categories. Furthermore, symmetry principles predict 
the existence of novel finer structures inside these normal subgroups 
forming feedforward and recurrent networks made of blocks of imprimitivity. 
These blocks constitute structures made of circulant matrices nested in a 
hierarchy of block-circulant matrices, whose functionality is understood in 
terms of neural processing filters responsible for fast processing of information.}

\end{abstract}

\maketitle

\section{Introduction}

There is growing consensus in present day complexity science that
functions of living networked systems are controlled by the structure
of interconnections between the network components
\cite{leibler,caldarelli,alon}. Under this assumption, the problem of
understanding how function emerges from structure~\cite{abbott} can be
cast in terms of the network structure itself, and this problem is,
fundamentally, of a theoretical nature.  Here we address this problem
by considering the connectome of the neural system of the nematode
{\it C. elegans}, a prototypical model connectome displaying complex
behavior
\cite{white,chalfie,bargmann2,chklovskii,varshney,zimmer,barabasi}.

Specifically, we show that the building blocks of the locomotion part
of the connectome are mathematically defined via its `{\it symmetry
  group}'~\cite{dixon}.  The implications of this result are two-fold.
First, we show that the symmetry group of locomotion circuits can be
broken down into a unique factorization as the {\it direct product} of
smaller `{\it normal subgroups}'~\cite[\S 1.6]{dixon}.  These normal
subgroups directly determine the separation of neurons into sectors.
The biological significance of this result 
%these sectors of neurons 
is measured by the fact that these sectors of neurons 
match known functional categories of the
connectome.  Second, we show that the sectors of neurons defined 
by the normal subgroups of the connectome can be further decomposed
into `{\it blocks of imprimitivity}'~\cite[\S1.5]{dixon} made of 
`{\it circulant}' matrices~\cite[\S3]{gray}.  
These circulant matrices are processing units encoding for fast signal
%processing 
filtering and oscillations in the locomotion function.
Figure 1a-g defines the group theoretical concepts
of permutation symmetry, normal subgroup, block of imprimitivity and
circulant matrix needed to understand the theoretical basis of the
structure-function relation in the connectome that we present here.

Our fundamental result is that symmetries of neural networks have a
direct biological meaning, which can be rigorously justified using the
mathematical formalism of symmetry groups. This formalism makes
possible to understand the significance of the structure-function
relationship: 
%the origin of the locomotion function in {\it C. elegans} 
the rationale behind the locomotion function in {\it C. elegans} 
is the existence of symmetries in the connectome which
uniquely assign neurons to functional categories defined through
the mechanism of factorization of the symmetry group. Therefore, the
structure-function relationship theoretically follows from a symmetry
principle. Although the specific form of the symmetry group
is different in different functions, the basic ideas and methods of
our formalism are the same and can be tested for any system. 
The symmetry group of a network has also a strong impact on the 
dynamics of the system. That is, it determines the synchronization 
of neurons belonging to the same orbit defined by the symmetry 
group~\cite{golubitsky,pecora1}. 
Furthermore, the form of synchrony determined by the symmetries of 
the connectivity structure is largely independent of the specific details 
of the neural dynamics. 
Seen globally, network symmetries may help to reveal the general 
principles underlying the mechanism of neural coding engraved in 
the connectome.

\section{Results}

\subsection{Neural connectome: symmetry groups}

The neuronal network of the hermaphrodite {\it C. elegans} contains
302 neurons, which are individually identifiable, and the wiring
diagram includes 890 gap junctions and 6393 chemical
synapses~\cite{varshney}.  The number of neurons across animals is
very consistent ~\cite{white,varshney}, while the gap-junctions and
chemical synapses are reproducible within 25\% variability from animal
to animal ~\cite{white,durbin,varshney,hall}.  Due to its
small size and relative completeness, the neural network of {\it
  C. elegans} has been a formidable model system to search for design
principles underpinning the structural organization and functionality
of neural networks
\cite{white,chalfie,bargmann2,chklovskii,varshney,zimmer,barabasi}.

We examine the forward and backward locomotion functions in {\it
  C. elegans} which have been well-characterized in the literature
\cite{white,chalfie,bargmann2,zimmer,barabasi,nematode,samuel,leifer}.
The locomotion is supported by two main functional classes of neurons
called (1) command interneurons and (2) motor neurons (in addition to
sensory neurons which are not studied here).  The backward locomotion
of the animal is supported by the activation of interneurons AVA, AVE,
and AVD \cite{nematode} and AIB and RIM \cite{zimmer}, and motor
neuron classes VA and DA. Similarly, forward locomotion is supported
by the activation of motor neuron classes VB and DB through the
interneuron classes AVB and
PVC~\cite{white,chalfie,nematode,zimmer,samuel} and RIB~\cite{zimmer}.
We use the most up-to-date connectome of gap-junctions and chemical
synapses from Ref.~\cite{varshney} to construct the neural circuits of
forward and backward locomotion (details in Supplementary Note 1. 
We represent the synaptic connectivity structure by the weighted 
adjacency matrix $A_{ij}\neq 0$ if neurons
$i$ connects to $j$, and $A_{ij}= 0$ otherwise. Gap-junctions are
undirected links, $A_{ij}=A_{ji}$, and chemical synapses are directed.

To explain the concept of symmetries and the procedure for finding the
symmetry group, we first consider the circuit comprising only the
interneurons connected via gap-junctions involved in the forward task
(Fig. 1a, adjacency matrix in Fig. 1b, weights on the links represent the number
of connections provided in~\cite{varshney}).  Later, we will see how
this circuit is integrated in the full connectome.

This sub-circuit contains $4!=24$ possible permutations of its 4
neurons. Out of these 24, only 8 are permutation symmetries as shown
in Fig. 1c.  A permutation symmetry, or
automorphism~\cite{dixon,pecora1,pecora2,other0}, is a transformation
defined as a permutation of neurons which preserves the connectivity
structure $A$ (see Supplementary Note 2
%SM Section II 
for detailed definition). 
This means that before and after the application of an
automorphism, the neurons are connected exactly to the same
neurons. Mathematically, if $P$ is an automorphism, then the permuted
adjacency matrix $PAP^{-1}$ is equal to the original one, $PAP^{-1} =
A$, or, equivalently, $P$ and $A$ commute with each other:
\begin{equation} 
[P,A] \equiv PA - AP = 0\ \ \ \iff\ \ \ P {\rm\ is\ a\ symmetry}\ .
\label{eq:commutatorPA}
\end{equation}

For instance, the permutation ABVL $\leftrightarrow$ RIBR and ABVR
$\leftrightarrow$ RIBL represented by $P_6$ in
Fig. 1c is an automorphism since it leaves the
connectivity intact.  The set of automorphisms forms the symmetry
group of the circuit, which, in this case, is the dihedral group ${\bf
  D_8}$, which is the group of symmetries of a square \cite{dixon}.
To be called a group of transformations, the transformations need to
satisfy four axioms: (1) the existence of an inverse in the group, (2)
the existence of an identity, (3) the associative law, and (4) the
composition law. In addition, if the transformations are commutative,
then the group is called abelian.

\subsection{Pseudosymmetries}

The study of the full locomotion circuit requires a generalization of
the notion of network symmetry, which we call `pseudosymmetry'.  The
concept of pseudosymmetry arises naturally from the observation that
connectomes vary from animal to animal, so no two worms will ever have
the same connectome~\cite{white,durbin,varshney,hall}. This variation
is estimated experimentally to be $25\%$ of the total connections from
worm to worm, as reported in~\cite{varshney} using data
from~\cite{white,durbin,hall}. We consider this variability across
individual connectomes as an intrinsic property consistent with
biological diversity and evolution. Furthermore, the number of
connections is subject to change from animal to animal through
plasticity, learning and memory~\cite{rankin}, so it cannot be
ignored.  On the other hand, while connectomes vary from animal to
animal, functions developed from them, such as forward and backward
locomotion, are barely distinguishable across different worms, and
still show some vestige of an ideal symmetry. In fact, the locomotion
function is preserved despite the 25\% variation in the connectomes.
Consequently, we expect that deviations from exact symmetries to be
relatively 'small'. Exact symmetries of the connectome should be
considered as an idealization, and we do not expect them to be
realized exactly.

Therefore, we consider pseudosymmetries of the connectome rather than
perfect symmetries.  Unlike perfect symmetries, defined by the abelian
commutator Equation~(\ref{eq:commutatorPA}) and shown in the circuit
of Fig. 1a, the definition of pseudosymmetry
depends on an additional parameter, a small number $\varepsilon> 0$.
This parameter quantifies the uncertainty in the connectivity
structure of the connectome due to natural variations across animals,
and, thus, we call it the {\it `uncertainty constant'} of the
connectome.  A pseudosymmetry is an approximate automorphism
$P_\varepsilon$, in the sense that the commutator
Equation~(\ref{eq:commutatorPA}) is replaced by a non-zero but small
$\varepsilon$-norm (detailed definition in Supplementary Note 3):
\begin{equation} 
||[P_{\varepsilon},A]||\ <\ \varepsilon
M\ \ \ \iff\ \ \ P_{\varepsilon} {\rm\ is\ a\ pseudosymmetry }\ ,
\label{eq:pseudocommutator}
\end{equation} 
where $M$ is the total number of network links including weights.
$P_\varepsilon$ approximates an exact symmetry in the ideal limit
$\varepsilon\to 0$.  The norm of the commutator, denoted as
$||[P_{\varepsilon}, A]||$, measures the number of links where
$P_{\varepsilon}$ and $A$ fail to commute given an upper limit
tolerance $\varepsilon$ in the fraction of links for the failure of
commutativity (a simple pseudosymmetry is exemplified in Fig. 1d).
The norm of the commutator in Equation~(\ref{eq:pseudocommutator}) is 
defined as the $L_1$ norm, denoted as $||[P,A]||$, and given by the 
following equation:
\begin{equation}
||[P,A]||=||P A-AP||=||A-P A P^{-1}||=\sum_{ij}|A_{ij}-A_{P(i)P(j)}|\ ,
\end{equation}
where the last equality follows from the fact that $P$ is an isometry
(i.e. $||A||=||PAP^{-1}||$ for any matrix $A$).  We see that this
definition of pseudosymmetry via a commutator resembles the
uncertainty principle in quantum mechanics and, thus, perfect
symmetries correspond to the `classical limit' of the
pseudosymmetries.

Equation~\eqref{eq:pseudocommutator} means that a pseudosymmetry must
preserve at least a fraction $(1 - \varepsilon)$ of network links.
The set of pseudosymmetries of the connectome contains not only the
symmetry group of the connectome $(\varepsilon=0)$ but it is augmented
by the permutations that are `almost' automorphisms.  We note that the
set of pseudosymmetries does not form a group by itself, since the
pseudosymmetries do not satisfy the composition law. For instance, two
pseudosymmetries (which by definition are below the threshold
$\varepsilon$) may be composed into a third pseudosymmetry that breaks
more than a fraction of $\varepsilon$ contacts, and, thus, does not
belong to the original set of pseudosymmetries, violating composition.

Knowledge of the pseudosymmetries is particularly useful for
understanding the robustness of functions under small perturbations of
the connectome. This property makes it analogous to the concept of
pseudospectrum which tells how much the spectrum of eigenvalues of a
matrix moves respect to small perturbations, see
\cite[$\S$2.8.1]{pseudospectrum}.  Simply put, if the set of
pseudosymmetries is clustered around the ideal symmetry group (i.e.,
the uncertainty constant is small), network functions are robust under
small perturbations. Conversely, if it is widely spread, then
functions are more likely to be lost under small perturbations.
Having all concepts at hand, we move to discuss the whole forward and
backward circuits and their symmetries, which, hereafter, are meant to
be pseudosymmetries, although we keep using the shorthand `symmetry'
for lexical convenience.

The forward gap-junction circuit is shown in
Fig. 2.  This circuit has permutation symmetries,
denoted as ${\bf F_{ gap}}$, most of which can be spotted by eye in
the layout displayed in the figure.  Figures 2a
and 2b display the real circuit, the adjacency
matrix and its pseudosymmetries (details of calculations in Supplementary Note 3). 
%SM Section III A).
The uncertainty constants $\varepsilon$ of these
pseudosymmetries are listed in Table~\ref{table:pseudo} and fall below
the upper experimental limit of 25\%. Thus, all pseudosymmetries have
biological significance. Figures 2c and 2d show an ideal circuit obtained 
by setting $\varepsilon=0$ compatible with the found pseudosymmetries 
%(see SM Section III A 
(see Supplementary Note 3 for details on how to obtain the ideal symmetric circuit).

\subsection{Symmetry group factorization into normal subgroups}

The crucial property of the symmetry group ${\bf F_{ gap}}$ is its
factorization into smaller {\it `normal subgroups'}. Its importance derives
from the fact that these normal subgroups match the known broad
functional categories of neurons involved in locomotion, such as
command interneurons, motor and touch neurons~\cite{wormatlas}.  A
`{\it subgroup}' ${\bf H}$ of a group ${\bf G}$ is a subset of
transformations of ${\bf G}$ which forms itself a group, i.e., the
transformations satisfy the four axioms of a group.

To understand what a normal subgroup is, we consider, for instance,
the automorphism that exchanges the motor neurons $\sigma:$ (VB2, DB3,
DB2, VB1) $\leftrightarrow$ (DB1, VB4, VB5, VB6) and forms (with the
identity) the dihedral group ${\bf D_1}$ (Figs. 2a
and 2b).  Importantly, this automorphism acts
independently only on neurons (VB2, DB3, DB2, VB1, DB1, VB4, VB5,
VB6), and leaves the rest of the neurons of the connectome intact.
Likewise, the automorphism $\tau:$ VB7 $\leftrightarrow$ VB3 forms
another group by itself, called the cyclic group of order 2, ${\bf
  C_2}$, and also acts independently on this set of neurons and not on
others.

The property of acting independently on a subset of neurons means that
${\bf D_1}$ (and ${\bf C_2}$) forms itself a smaller group, called a
`{\it normal subgroup}' inside the full symmetry group ${\bf F_{
    gap}}$.  More formally, a subgroup ${\bf H}$ is said to be normal
in a group ${\bf G}$ if and only if ${\bf H}$ commutes with every
element $g\in {\bf G}$, i.e., $[g, {\bf H}] = g{\bf H} - {\bf H}g=0$
The formal definition of subgroup and normal subgroup are explained in 
Supplementary Note 4, see~\cite[\S1.6]{dixon}.
 
This property implies that the group ${\bf F_{ gap}}$ can be
factorized in a unique way as a direct product of its two normal
subgroups as: ${\bf D_1} \times {\bf C_2}$ (definition of
factorization of a group in Supplementary Note 4, 
%SM Section~\ref{sec:factorization},
see~\cite[\S1]{dixon}).  The significance of the normal subgroup is
that the normal transformations identify a unique and non-overlaping
subset of neurons that are moved by each normal subgroup. This set of
neurons are called the `{\it sector}' associated with the normal
subgroup.  Since each normal subgroup acts only on an independent
sector, the factorization of groups into normal subgroups leads also
to a partition of neurons into unique disjoint sectors. 

In simple terms, this means that when an automorphism in a normal
subgroup is applied to the network, only the neurons in the sector of
the normal subgroup are permuted, while the rest of the neurons that
are outside the sector are not affected.  Thus, we say that the normal
subgroup automorphisms act only on the neurons belonging to its sector
providing a unique separation and classification of the neurons and
the associated factorization of the symmetry group. This factorization
is mathematically analogous to the unique factorization of natural
numbers into primes, and this notion is extended to group theory for
those finite groups that can be factorized into `prime' normal
subgroups, as it is the case of the connectome.

The  symmetry group ${\bf F_{ gap}}$ is factorized as
a direct product of $6$ normal subgroups as:
\begin{equation}
{\bf F_{ gap}} =\ [{\bf C_2}\times {\bf C_2}]\ \times\ 
[{\bf S_5}\times {\bf D_1} \times {\bf C_2}\times {\bf C_2}] \ .
\label{eq:Fgap}
\end{equation}
Each subgroup acts on a non-overlapping independent sector of neurons
as indicated in Fig. 2 (see also Supplementary Note 5). 
%SM Section V A).
Table~\ref{table:pseudo} lists the uncertainty constant for each subgroup 
of pseudosymmetries indicating that all $\varepsilon$ are small and below 
the experimental upper limit 25\%~\cite{white,durbin,hall,varshney}.

The factorization of the symmetry group ${\bf F_{ gap}}$ in
Equation~\eqref{eq:Fgap} is significant because it determines a
partition of the circuit into sectors that match specific categories 
of neurons~\cite{wormatlas}.
To define the functional categories or classes of neurons we follow 
the literature where functions have been determined experimentally 
and compiled at the WormAtlas~\cite{wormatlas}. Broad functional 
categories of neurons are provided at 
\url{http://www.wormatlas.org/hermaphrodite/nervous/Neuroframeset.html},
Chapter 2.2. A classification for every neuron into four broad neuron 
categories follows: (1) motor neurons, (2) sensory neurons, (3) interneurons, 
and (4) polymodal neurons. A function is assigned to each neuron based on 
this experimental classification into neuron categories. This classification is 
displayed in Supplementary Table 1 and Supplementary Table 2, 
and discussed in Supplementary Note 6. 
These categories represent the ground truth to test the predictions of our theory. 

Specifically, the factor $[{\bf C_2}\times {\bf C_2}]$ corresponds to
the command interneuron category and comprises command interneurons
which drive the forward locomotion: AVBL, ABVR and RIBL, RIBR. The
entire motor class is associated to an entire independent factor
$[{\bf S_5}\times {\bf D_1}\times {\bf C_2}\times {\bf C_2}]$, and
includes all motor neurons innervating the muscle cells responsible
for the undulatory motion of {\it C. elegans}.  

Applying the same symmetry procedure, we find that all
forward/backward gap-junction/chemical synapse circuits form symmetry
groups, and these groups can be factorized into normal subgroups in
the same way as in Equation~(\ref{eq:Fgap}) (see
Fig. 3 and Fig. 4 and Supplementary Note 5 for details). 
The correspondence between
neuron sectors from group theory and known categories of neurons
occurs consistently across all circuits, and further includes a
subgroup related to touch sensitivity~\cite{chalfie} in the forward
(PVCL, PVCR, Fig. 4a) and backward (AVDL, AVDR,
Fig. 4c) chemical synaptic circuits. The full list
of sectors, normal subgroups and uncertainty constants of the
pseudosymmetries are provided in Table~\ref{table:pseudo}.

The normal subgroups partition the connectome into non-overlapping
functional sector of neurons, thus realizing the segregation of
function.  At the same time, the sectors remain connected in the
connectome without breaking the symmetries, thus fulfilling the
integration of function into a globally connected network. Thus, the
symmetric subgroup organization of the connectome provides an elegant
solution to the conundrum of functional specialization in the presence
of a global integration of information necessary for efficient
coherent function~\cite{tononi}, a profound issue in neuroscience

While group factorization can distinguish different classes of
neurons, this distinction may also be seen in some cases by directly
looking at the adjacency matrix: for instance in
Fig. 2b the AVB interneurons are heavily connected
hub neurons which could be, in principle, also distinguished by any
connectivity measure. That is, the neurons AVBL and AVBR are hubs with
large degree $k=18$ and are easily distinguished from the rest of the
neurons which have generally smaller degree.  However, in general,
having the same degree does not imply that the neurons belong to the
same subgroup. Thus, the connectivity measure alone may not fully
capture the symmetry groups that we find.

For instance, neurons can be in the same sector subgroup and at the
same time could, in principle, have different degree. This situation
is seen for example in the neurons of the forward motor sector
subgroup ${\bf D_1}$ in Fig. 2a
and 2c. In the circuit of
Fig. 2c, the neurons in ${\bf D_1}$ have different
degree: VB5, DB2 VB4 and DB3 with $k=5$, VB1 and VB6 with $k=3$, VB2
and DB1 with $k=8$. Thus, even though these neurons have different
degree, they belong to the same subgroup and functional class: the
motor sector subgroup ${\bf D_1}$. In general, the degree alone is not
enough to separate the neurons in subgroups and known classes.

Furthermore, Fig. 2b shows that the pair (VB8,
VB9) has the same connectivity as the pair (RIBL, RIBR), and thus they
could be classified in the same category as either motor neuron (with
VB) or interneurons (with RIB).  If we consider the neurons unweighted
they merge into the same subgroup and they should perform the same
function. However, considering the weights, there is an asymmetry,
since both, VB9 and VB8 have 6 and 7 connections to AVBL and AVBR
respectively, while RIBL and RIBR have one connection each to both
neurons (see Supplementary Fig. 3). Indeed, the WormAtlas
classifies RIBL and RIBR as interneurons~\cite{wormatlas}, thus, we
classify these pairs of neurons in different classes. The asymmetry in
the weights might the reason why the experiments compiled at the
WormAtlas find that these two set of neurons may work in different
categories: motor and interneuron.  In general, it is possible for a
neuron to be involved in multiple functions.  The case of polymodal
neurons can be treated theoretically by generalizing the
direct-product factorization to semidirect-product factorization of
normal and non-normal subgroups. Semi-direct product factorization
could capture overlapping sectors of neurons and multi-functionality
which are more prevalent across the connectome beyond locomotion.

\subsection{Statistical significance of the symmetry subgroups}
\label{significance}

To establish the degree to which the symmetries of the locomotion
sub-circuits are statistically significant, we compare the symmetry
subgroups against control random sub-circuits. Indeed, a high enough
value of $\epsilon$ would yield an approximate symmetric version of
any arbitrary circuit: a fully random non-symmetric connectome implies
$\epsilon = 1$, and a perfect symmetric one $\epsilon = 0 $. In
between, all networks can be classified by their
$\epsilon$-value. Thus, it is important that not only $\epsilon$ be
smaller than the experimental variability $\epsilon<25\%$, but also be
statistically significant. Statistical metrics to evaluate the
symmetries are p-value statistical tests to compare results with a
randomized null model preserving the degree sequence.
Specifically, the p-value of a pseudo-symmetry subgroup $G_{\epsilon}$
is defined as the probability to find a subgroup $G_{\epsilon^*}$ with
$\epsilon^*\leq\epsilon$ in a randomized circuit with the same degree
sequence as the real circuit. The results of the p-values are
summarized in Table~\ref{table:pseudo} for each subgroup, showing that
pseudosymmetry subgroups are, indeed, statistically significant.

\subsection{Comparison with other methods to find functional modules}
\label{sec:modularity}

It is interesting to compare the functional partition obtained by the
symmetries of the connectome with typical modularity detection
algorithms which are widely used to identify functional modules in
biological networks~\cite{newman}. Indeed, there is a large body of
work which examines the connectivity of biological networks to
algorithmically classify the constituent neurons into modules and
compare those modules to known classifications. Therefore, below 
we investigate how symmetry detected sectors compare to existing 
algorithms such as modularity and community detection, and other 
centrality measures. 

We run the Louvain community detection algorithm~\cite{blondel} on the
forward and backward circuit and find the modular partition seen in
Fig. 5. We find that modules identified by the
Louvain algorithm do not generally capture the functional modules
identified by symmetry subgroups, nor the experimental classification
into neural functions.

Typically, the modularity algorithm assigns to the same functional 
module a hub-like interneuron AVBR together with its connected 
neurons in the motor sector (see Fig. 5a), since 
these neurons are all highly connected. Thus, the modularity algorithm 
will typically mix the interneuron and motor sectors. 
Symmetry factorization into normal subgroups, on the other hand, correctly
classifies AVBL and AVBR separately from the VB and DB neurons in the
motor sector, even though these sectors are well connected. 
Similar results are obtained when we use other network centralities: 
Fig. 5b and 5d show the modules
obtained by ranking neurons according to eigenvector centrality. We 
find that such centrality measure does not capture the partition into
symmetry sectors nor the functional classes.  

\subsection{A recurrent and feedforward neural network made of blocks 
of imprimitivity and circulant matrices}

The data analyzed so far indicate that there is still a more refined
structure inside the broad functional categories of motor, command and
touch, that requires further exploration. For instance, the motor
class of forward gap junctions (Fig. 2) consists of
4 different normal subgroups: $[{\bf S_5}\times {\bf D_1}\times {\bf
    C_2}\times {\bf C_2}]$.  Next, we show that the functionality of
this finer structure can be systematically obtained through a more
refined group theoretical concept of {\it `block of imprimitivity'}
\cite[$\S 3$]{dixon}, which identifies the fundamental processing
units of the connectome and naturally leads to a novel functionality
in terms of mechanism of neural coding.

A block of imprimitivity is a set of neurons that, under the action of
the automorphisms of a subgroup, is completely mapped onto itself or
it is mapped onto a completely disjoint set of neurons (formal
definition of block of imprimitivity in Supplementary Note 7, 
see~\cite[$\S 1.5$]{dixon}, and
Fig. 1f). For instance, consider the subgroup ${\bf
  D_1}$ of the forward gap-junction circuit
(Fig. 2) which consists of the automorphism
$\sigma\in {\bf D_1}$ which acts on the sector (VB2, DB3, DB2, VB1,
DB1, VB4, VB5, VB6). The subset of neurons highlighted in green in
Fig. 2c, ${\mathcal B_1}=$ (VB2, DB3, DB2, VB1),
forms a block of imprimitivity since $\sigma$ moves this set into a
different one, highlighted in black, ${\mathcal B_2}=$ (DB1, VB4, VB5,
VB6), which is the other block of imprimitivity of the sector and a
conjugate block of ${\mathcal B_1}$. 
These two blocks form the so-called {\it system of imprimitivity}, 
a fundamental concept in group theory~\cite{dixon,weinberg}.
The other normal subgroups of the forward circuit do not have a 
nontrivial block system of imprimitivity, hence they are said to be 
{\it primitive} (Supplementary Note 7, \cite[\S3]{dixon}).

The resulting block partition of each adjacency matrix is shown in
Figs. 2d, 3d, 4b and 4d. These systems of imprimitivity identify new 
functionalities in each locomotion circuit.
Specifically, we find that the system of imprimitivity of each
locomotion circuit is formed by blocks represented by {\it circulant}
matrices~\cite{gray}.  A circulant matrix is a square matrix where
each row is a cycle shift to the right of the row above it, and
wrapped around~\cite[\S3]{gray} (see Methods Section
for definition). In alignment with pseudosymmetries, the circulant matrices
are interpreted as pseudocirculant matrices of the real circuit. A
pseudocirculant matrix differs from a circulant matrix by a fraction
$\varepsilon$ of their links. 
We note that this partition into blocks of imprimitivity is not 
unique. For instance, another possible block system corresponds 
to a partition made by the orbits. 

Circulant matrices are well-known in the field of digital signal
processing, recurrent and feedforward neural networks~\cite{abbott}
and cryptography, and are widely used as efficient linear filters to
solve a variety of tasks in digital image processing, most notably as
edge-detection and signal compression~\cite{abbott,DSP}, but also in
tracking~\cite{tracking}, voice recognition, and computer vision~\cite{lecun}.
Circulant matrices are the kernels of discrete convolutions and are
used in discrete Fourier transform to solve efficiently systems of
linear equations in nearly linear time~\cite{gray} that significantly
speed up the $O(N^3)$ arithmetic complexity of Gaussian elimination.

We find different types of circulant matrices in the connectome which
are, in turn, nested into larger block-circulant matrices (see
definitions in Fig. 1b and Fig. 2d and Methods Section).
Two circulant matrices occur consistently in all locomotion circuits and
act as a `high-pass' filter:
\begin{equation}
  {\mathcal H}= {\rm circ}(0,
1)=\begin{bmatrix} 0&&1\\ 1&&0
\end{bmatrix},
\label{highpass}
\end{equation}
and a `low-pass' filter:
\begin{equation}
  {\mathcal L} = {\rm circ}(1, 1) =
\begin{bmatrix} 1& &1\\ 1&&1
\end{bmatrix}.
\label{lowpass}
\end{equation}
The third type of
circulant matrix represents a 4-cycle permutation:
\begin{equation}
  {\mathcal F} = {\rm
  circ}(0, 1, 0, 1) = \begin{bmatrix} 0 && 1 && 0 && 1\\ 1 && 0
    && 1 &&
    0\\ 0 && 1 && 0 && 1\\ 1 && 0 && 1 &
    &0\\
  \end{bmatrix},
\end{equation}
and  acts on the
blocks of imprimitivity ${\mathcal B_1}$ and ${\mathcal B_2}$ in the
motor sector of the forward gap-junction circuit
(Figs. 2c and 2d).
Intuitively, each circulant matrix represents a cycle embedded in the
subgroup sector as seen in Fig. 2c for ${\mathcal
  B_1}:$ VB2 $\to$ DB3 $\to$ DB2 $\to$ VB1 $\to$ VB2.
In the same figure we see the 4-cycle of the conjugate block
${\mathcal B_2}$.

The $2\times 2$ circulant matrix ${\mathcal H}$ in Equation~\eqref{highpass} 
is quite ubiquitous and corresponds to a 2-cycle (or transposition). 
For instance, the 2-cycle VB8 $\to$ VB9 $\to$ VB8 in the forward 
gap junction circuit Fig. 2c forms a circulant matrix 
of the form given by ${\mathcal H}$. This is also a block of imprimitivity, 
since this block is the only one inside the subgroup ${\bf C_2}$.  
Subgroup ${\bf S_5}$ also forms a circulant matrix, although a trivial one 
in this case since all its elements are zero.

It is interesting to see that the circulant matrices are nested into
an structure of block-circulant matrices (see Methods Section
for definition), suggesting a hierarchical
organization of building blocks in the connectome.  Typical
block-circulant matrices are of the form~\cite{gray}:
\begin{equation}
\mathcal{BC}\ = \  {\rm bcirc}( {\mathcal H},  {\mathcal L} ) \ =\  
\begin{bmatrix}
   {\mathcal H}  &  & {\mathcal L} \\
    {\mathcal L} &  & {\mathcal H}   
\end{bmatrix}
\ =\  
\begin{bmatrix}
    0 &  & 1 & & 1 & & 1 \\
    1 &  & 0 & & 1 & & 1 \\
    1 &  & 1 & & 0 & & 1 \\
    1 &  & 1 & & 1 & & 0 
\end{bmatrix}
\ .
\label{eq:bcirculant}
\end{equation} 
For instance, this block-circulant matrix appears in the command
sector of the forward gap junction circuit between the neurons RIBL,
RIBR, AVBL, AVBR. This is seen in Fig. 1b and also
in Fig. 2d. It is interesting to note that when we
analyze the group structure of the interneuron only circuit of gap
junctions, then we find the group structure shown in
Fig. 1b. When we integrate this circuit in the full
forward circuit, then this group becomes a system of imprimitivity
shown as ${\mathcal B_6}$ and ${\mathcal B_7}$ in
Fig. 2d.  This is a block-circulant matrix made
itself by circulant matrices forming a nested hierarchical
structure. This hierarchical nestedness is repeated across all the
connectome.

A block-circulant structure is formed by the imprimitive
blocks ${\mathcal B_1}$ and ${\mathcal B_2}$ in the same forward gap
junction circuit, Fig. 2d. In this case, we have:
\begin{equation}
A_1={\mathcal F} = {\rm circ}(0, 1, 0, 1) = 
 \begin{bmatrix}
    0 &  & 1 & & 0 & & 1 \\
    1 &  & 0 & & 1 & & 0 \\
    0 &  & 1 & & 0 & & 1 \\
    1 &  & 0 & & 1 & & 0 
\end{bmatrix} , \, \, \, \, \, \, \, \,
{\rm and} \, \, \, \, \, \, \, \,
A_2 = 
 \begin{bmatrix}
    0 &  & 0 & & 0 & & 1 \\
    0 &  & 0 & & 0 & & 0 \\
    0 &  & 1 & & 0 & & 0 \\
    0 &  & 0 & & 0 & & 0 
\end{bmatrix} ,
\end{equation}
 and both  ${\mathcal B_1}$ and ${\mathcal B_2}$
combine into a block-circulant matrix of the form: 
\begin{equation}
{\mathcal BC} =
{\rm bcirc}(A_1, A_2).
\end{equation}
Also, ${\mathcal B_6}$ and ${\mathcal B_7}$ in the backward gap
junction circuit of Fig. 3d composed of neurons
AIBL, AIBR, RIML, RIMR form a block-circulant matrix 
\begin{equation}
{\mathcal BC} =
{\rm bcirc}(A_1, A_2)
\end{equation}
 with 
\begin{equation}
A_1 = {\rm circ}(0, 0) \, , \, \, \, \, \, \, \, \, {\rm and} \, \, \, \,
\, \, \, \, A_2 \ =\ {\rm circ}(1, 0) =
\begin{bmatrix}
    1 &  & 0 \\
    0 &  & 1   
\end{bmatrix}
\ .
\end{equation}

These results suggest that we can think of the connectome as a
feedforward network made of interneurons that feeds a recurrent
network in the motor system~\cite{abbott} made of a system of sensing
operators, each represented by an imprimitive block with a circulant
structure.  Such a feed-forward and recurrent network architecture is
universally seen across many neural systems and it is used as a model
of the receptive fields in the primary visual cortex
\cite{abbott}. Such a system can be modeled by a feedforward matrix
${\bf W}$ and a recurrent network ${\bf M}$ processing the input
activity ${\bf u}$ to the output ${\bf v}$ as a linear filter, see
Dayan \& Abbott \cite{abbott}:
\begin{equation}
  \tau \frac{d {\bf v}}{dt} = - {\bf v} + {\bf M} {\bf v} + {\bf W}
       {\bf u},
       \label{linear}
\end{equation}
where $\tau$ is a time characteristic. The crucial property of this
system is that the matrix ${\bf M}$ contains loops in the network.

For instance, in the case of the gap junction forward circuit
(Fig. 2), the AVBL and AVBR interneurons act as the
input layer ${\bf u} = (u_{\rm ABVL}, u_{\rm ABVR})^T$ which is first
processed by the feedforward matrix represented by a fully connected
matrix:
\begin{equation}
  {\bf W} = 
 \begin{bmatrix}
    1 &  & 1 \\
    1 &  & 1 \\
    1 &  & 1 \\
    1 &  & 1
\end{bmatrix} ,
\end{equation}
whose output is then processed by the recurrent network in the motor
sector by, for instance, processing the signal in the motor neurons of
the imprimitivity block ${\mathcal B_1}$, ${\bf v} = (v_{\rm VB2},
v_{\rm DB3}, v_{\rm DB2}, v_{\rm VB1})^T$ by the recurrent circulant
matrix ${\bf M} = {\mathcal F} = {\rm circ}(0,1,0,1)$ from
Equation~(\ref{eq:bcirculant}).  The same signal processing occurs in the
feedforward and recurrent network formed by the conjugate motor
imprimitive block ${\mathcal B_2}$.  Similar structure is seeing in
the backward circuit Fig. 3 with AVAL-AVAR
feed-forwarding information into the recurrent circulant blocks in the
motor sector.  The chemical circuits also contain such a feed-forward
and recurrent structure: PVCL-PVCR feeds the forward motor circulant
blocks (Fig. 4a) and AVE-AVD-AVA feed the backward
motor circulant blocks (Fig. 4c).

Using the language of signal processing in computational neuroscience,
these recurrent networks are analogous to the core of receptive fields
that process information in the visual cortex, see Dayan \& Abbott
\cite{abbott}. For instance a widely used filter in signal processing
is the edge-detector \cite{abbott,DSP} which employs a circulant
matrix defined by ${\bf M} = {\rm circ}( 0, 1, -1, 0, \cdots, 0)$ to
compute a `derivative' of the spatial signal and detect sharp edges
\cite{abbott}. Another typical computation is performed by a circulant
matrix ${\bf M} = {\rm circ}( 0, 1, -2, 1, 0, \cdots, 0)$ to represent
a second derivative of the signal, and so on.

In the case of the connectome, one possible interpretation of 
the purpose of the found circulant filters is to separate one band of frequencies 
from another and perform signal compression.
The high-pass filter ${\mathcal H}$ is used to block the
low frequency content of the neural signal, while the low-pass filter
eliminates the high frequencies.  The ${\mathcal F}$ matrix is a
translational invariant filter to sample the signal as a way of
reducing the size of the signal (compression) without overly reducing
its information content to process the undulatory motion of locomotion
according to its eigenvalues.

Roughly speaking, the filter $\mathcal{H}$ measures the
self-similarity on either side of the center point and the output will
be maximal when each the two points are equal to each other.  The
filter $\mathcal{F}$ operates on the inputs of the imprimitive systems
of the forward circuit.
The fact that this matrix appears only in the forward circuit suggests
that it might be an important controller in the undulatory
motion. This can be seen from the eigenvalues $\lambda_i$ of this
circulant matrix and their eigenvectors ${\bf v_i}$:

\begin{equation}
\begin{matrix}
  \lambda_1 &=& -2,& \,\, \,\, \,\, & {\bf v_1} &=& \frac{1}{2}
  (-1, 1, -1, 1),  \\
  \lambda_2 &=& 2, &\,\, \,\, \,\, &  {\bf v_2}  &=& \frac{1}{2}
  (1, 1, 1, 1),  \\
  \lambda_3 &=& 0, &\,\, \,\, \,\, &  {\bf v_3}  &=& \frac{1}{\sqrt{2}}
  (0, -1, 0, 1),  \\
  \lambda_4& =& 0,& \,\, \,\, \,\, &  {\bf v_4} &=& \frac{1}{\sqrt{2}}
  (-1, 0, 1, 0), 
\end{matrix} \,
\label{eigenvalues}
\end{equation}
which determine the solution of Equation~(\ref{linear}) \cite{abbott} and
act by filtering out two modes and allow oscillations between
$\lambda_1$ and $\lambda_2$.  Thus, the circulant blocks act as
information processing units in the recurrent network that are
basically filters to perform specific signal processing operations
(see Supplementary Note 8). 

The association of the circulant processing units with the blocks of
imprimitivity completes the operational definition of the locomotor
function determined by the decomposition properties of its symmetry
group, and in turn, unveil and classify hitherto hidden mechanisms of
the neural code. The existence of the predicted blocks can be directly
tested in future experiments by measuring how the imprimitive blocks
process the neural signal in real time according to their circulant
filters.

\section{Discussion}

Overall, the structure-function relation in the connectome can be seen
as a refining process of nested symmetry building blocks.  The primary
building blocks are defined through the mechanism of direct product
factorization of normal subgroups and provide a rigorous
characterization of the network connectivity structure, and a simple
interpretation of its major functions into neural classes.  These
major sectors are comprised of secondary topological structures
involved in signal processing which refine the primary normal
subgroups into irreducible blocks of imprimitivity.

  The factorization of the symmetry groups of the
  connectome has its analogy with integers and primes as every integer
  can be factorized into a unique product of prime numbers as stated
  in the fundamental theorem of arithmetic. This factorization is also
  analogous to that of the Standard Model of particle
  physics~\cite{weinberg}. In theoretical physics, automorphisms
  describe the symmetries of elementary particles and
  forces~\cite{weinberg,georgi}, as well as atoms, molecules and
  phases of matter~\cite{landau}. For example, fundamental forces in
  particle physics are based on symmetry principles incorporated
  through a description of the gauge symmetry group of the Lagrangian
  factorized into three subgroups as ${\rm U}(1) \times {\rm SU}(2)\times {\rm SU}(3)$,
  where ${\rm SU}(N)$ is the special unitary group of $N\times N$ unitary
  matrices with determinant 1, and ${\rm U}(1)$ is the group consisting of
  all complex numbers with absolute value 1. In this case, each
  subgroup determines a different force, namely the electroweak and
  strong forces, and the generators of these symmetry subgroups are
  the particles. Analogously, the functions of locomotion are based on
  the symmetries of the connectome through the symmetry group which is
  factorized in general as ${\rm T}\times {\rm C}\times {\rm M}$ where
  each symmetry subgroup determines a different function. For instance, 
  the symmetry group of the chemical forward circuit splits as: 
  ${\bf F_{ch}} = {\mathbb T}_{\bf F_{ch}} \times {\mathbb C}_{\bf F_{ch}} \times {\mathbb M}_{\bf F_{ ch}}$.  

In a milestone in the history of mathematics, all finite simple groups
have been discovered and classified into 3 major classes: cyclic,
alternating or Lie type plus 26 extra classes of rare sporadic groups
\cite{gorenstein}.  Out this variety, the locomotion connectome
contains only cyclic groups.  It would be fascinating to discover
other naturally occurring simple groups for other functions in
different biological networks. Results presented elsewhere indicate
that symmetries extend to the full connectome and also to genetic
networks \cite{genetics}, and they are naturally related to neural
synchronization.  Thus, the principle of symmetry provides a rigorous
mathematical characterization of the structural and functional
organization of connectomes down to their information-processing
units. This hierarchical symmetric architecture may also serve as
guidance to design more efficient artificial neural networks inspired
by natural systems.

\clearpage

\section{Methods}
\label{circulant}
{\bf Circulant and block-circulant matrices in digital signal
  processing and the connectome}

We find that the system of imprimitivity of the locomotion circuits is
comprised of a specific type of blocks, which are represented by
circulant matrices \cite{gray},
\url{https://en.wikipedia.org/wiki/Circulant_matrix}.

It is worth noting that there is {\it a priori} no reason for the
occurrence of this specific type of matrices in the system of
imprimitivity.  That is, a symmetry group may have a system of
imprimitivity that is not composed of circulant matrices. Thus, there
are two independent results: first, the connectome is broken down into
a system of imprimitivity.  Second, the imprimitive blocks have the
shape of circulant matrices and block circulant matrices. 

A circulant matrix $P_{\ell}$ of order $\ell$ is a square matrix of
the form \cite{gray}:
\begin{equation} 
 P_{\ell}\ =\ {\rm circ}(c_1, c_2, \ldots, c_\ell) \ = \  
\begin{bmatrix}
    c_1 & c_2 & c_3 & & \ldots   & & c_\ell  \\
     c_\ell  & c_1 & c_2 & & \ldots  & & c_{\ell-1} \\
     \cdot  & \cdot  &    & &  & & \cdot \\
     \cdot  & \cdot  &    & &  & & \cdot \\
   c_2  & c_3 & c_4   & &  \ldots  & & c_1 
\end{bmatrix} \ .
\label{eq:PL}
\end{equation}
The elements of each row are the same as those from the previous row,
but are shifted one position to the right and wrapped around.
The circulant matrix is thus determined by the first row or column and
therefore it is denoted by \cite{gray}: $ P_{\ell}\ =\ {\rm circ}(c_1,
c_2, \ldots, c_\ell)$.

We also find block-circulant matrices in the connectome which are
defined as follows. Block-circulant matrices are an extension of
circulant matrices where the elements $c_i$ are now replaced by
matrices themself $A_i$. Let $A_1, A_2, \dots, A_m$ be square
matrices of order $n$. A block-circulant matrix of order $m n$ is the
form \cite{gray}:

\begin{equation} 
 \mathcal{BC}\ =\ {\rm bcirc}(A_1, A_2, \ldots, A_m) \ = \  
\begin{bmatrix}
    A_1 & A_2 & A_3& \ldots   & & A_m  \\
     A_m  & A_1 & A_2 & \ldots  & & A_{m-1} \\
     \cdot  & \cdot  & &  & & \cdot \\
     \cdot  & \cdot  & &  & & \cdot \\
   A_2  & A_3 & A_4 &  \ldots  & & A_1 
\end{bmatrix}\ ,
\label{eq:BC}
\end{equation}
and when $n=1$, the block-circulant becomes a circulant matrix. The
matrices $A_i$ may not need to be necessarily circulant. However, the
connectome presents only circulant matrices as $A_i$, thus creating a
hierarchical nested structure of circulant blocks made of circulant
matrices themself.

The graph that results from a circulant matrix is called a circulant
graph, \url{https://en.wikipedia.org/wiki/Circulant_graph}.  Circulant
matrices are determined by the first row and every row is the cyclic
shift of the row above it.  A circulant matrix is a special kind of
Toeplitz matrix with the additional property that $c_i = c_{i + \ell}$
\cite{gray}.

Repeated application of $P_{\ell}$ on itself generates an abelian
group called {\bf cyclic group} of order ${\ell}$, denoted as ${\bf
  C_{\ell}}$. Moreover, any subgroup of ${\bf C_{\ell}}$ is also
cyclic.  The important point is that whenever the symmetry group of a
network contains a circulant permutation matrix like $P_{\ell}$ in
Equation~\eqref{eq:PL}, then the adjacency matrix $A$, or a piece of it,
inherits from $P_{\ell}$ the same circulant structure.

In the locomotion neural circuits studied in this work, we find 3 types of
circulant matrices: ${\mathcal H}, {\mathcal L},$ and ${\mathcal F}$. In the
language of signal processing, the matrix $\mathcal H$ is a spatial
high-pass filter, used to block the low frequency content of the
signal; and $\mathcal L$ is a spatial low-pass filter, which instead
eliminates the high frequencies.  These are the two most common linear
filters used in image processing.  The filter $\mathcal F$ is the
kernel of the fast Fourier transform (see Supplementary Note 8). 
It can be thought as a translational invariant
filter to sample the signal as a way of reducing the size of the
signal without overly reducing its information content.  While
$\mathcal H$ and $\mathcal L$ appear consistently across all circuits,
the circulant $\mathcal F$ matrix occurs only in the forward
gap-junction circuit.  The low pass filter selects the `bulk' of the
information, while the `high-pass' picks out finer details.

This structure shows how the connectome acts as a signal processing
network within a hierarchical structure that starts at the symmetry
group level, which is then broken down into subgroups and further
broken into the system of imprimitivity which represents the
irreducible building blocks.

\vspace{1cm}

{\bf Data availability:} Connectome data are available in the public
domain at~\url{http://www.wormatlas.org} and codes
at~\url{http://www.kcorelab.org} and \url{http://github.com/Makselab}.

\vspace{1cm}

{\bf Acknowledgments:} We are grateful to L. Parra, A. Holodny, and
I. Leifer for discussions.  This work was supported by grants from
NIH-NIBIB R01EB022720, NIH-NCI U54CA137788 / U54CA132378 and NSF-IIS
1515022.

{\bf Authors contributions:} F.M. and H.A.M. contributed equally to this
work.

{\bf Competing interests:} Authors have no competing interests.

\vspace{1cm}

FIG. 1. {\bf Group theoretical definitions:
  automorphism, symmetry groups, pseudosymmetries, normal subgroups,
  and blocks of imprimitivity}.  {\bf (a) } Circuit made of
gap-junction and only interneurons in the forward locomotion used to
define an automorphism.  These are permutation symmetries that leave
the adjacency structure invariant.  These symmetries then convert to a
system of imprimitivity when we integrate the circuit into the full
locomotion connectome. Nodes represent neurons and weighted links
represent the number of gap-junctions connections between neurons from
Ref.~\cite{varshney}.  {\bf (b)} Adjacency matrix of the circuit in
{\bf (a)}. This matrix is composed of circulant matrices: a high-pass
filter ${\mathcal H}={\rm circ}(0, 1)$ in the diagonal and an
off-diagonal low-pass filter ${\mathcal L} = {\rm circ}(1, 1)$. The
full $4\times 4$ matrix forms a block-circulant matrix ${\mathcal BC}
= {\rm bcirc}({\mathcal H}, {\mathcal L})$ \cite{gray} (see Methods Section
for definitions).  {\bf (c)} Symmetry group of the
circuit shown in {\bf (a)}, called dihedral group ${\bf D_8}$,
comprises 8 automorphisms out of the 4!  = 24 possible permutations of
neurons. We show each permutation matrix $P$ of each automorphism.
{\bf (d)} Pseudosymmetries capture inherent variabilities in the
connectome from animal to animal. An example pseudosymmetry is shown
$P_\varepsilon=$ DB5 $\leftrightarrow$ DB6 that breaks one link to
AVBR over 18 total weighted links, giving $\varepsilon=1/18=5.5\%$.
{\bf (e)} Definition of normal subgroup. A subgroup ${\bf H}$ is said
to be normal in a group ${\bf G}$ if and only if ${\bf H}$ commutes
with every element $g\in{\bf G}$, that is: $[g, {\bf H}] = g{\bf H} -
{\bf H}g = 0$ (see Supplementary Note 4 for a detailed explanation).  
{\bf (f)} Definition of blocks of imprimitivity and
system of imprimitivity.  Simply put, a set of nodes is called a block
(of imprimitivity) if all nodes in this set always `move together'
under any automorphism of the symmetry group. A set of blocks with
such a property is thus called a system of imprimitivity (see 
Supplementary Note 7 for a formal definition).  {\bf (g)}
Definition of circulant matrix and circular convolution. Matrix
$\mathcal{F}$ appears in the forward gap-junction locomotion circuit
and is called a circulant matrix. This matrix has a peculiar pattern
where each row is a shift to the right by one entry of the previous
row.  Multiplication of $\mathcal{F}$ by a vector $\bf x$ gives rise
to a famous operation called a circular convolution, which is used in
many applications, ranging from digital signal processing, image
compression, and cryptography to number theory, theoretical physics
and engineering, often in connection with discrete and fast Fourier
transforms, as explained in Supplementary Note 8. 

\bigskip

FIG. 2. {\bf Symmetry group ${\bf F_{gap}}$ of the
  forward gap-junction circuit.}  {\bf (a)} Circuit from
Ref. \cite{varshney}. Pseudosymmetries $P_\varepsilon$ act on distinct
sectors of neurons indicated by different colors that lead to direct
product factorization of the symmetry group into normal subgroups.
The normal subgroups sectors of neurons match the broad classification
of command interneurons and motor neurons from the Wormatlas
\cite{wormatlas}.  {\bf (b)} Adjacency matrix of {\bf (a)} showing the
normal subgroup structure and its matching with broad neuronal
classes. {\bf (c)} Idealization of the circuit obtained from {\bf (a)}
by $\varepsilon\to 0$ leading to perfect symmetries (see Supplementary Note 3). 
%SM Section III A).
We highlight the two 4-cycles across $\mathcal B_1$:
VB2 $\to$ DB3 $\to$ DB2 $\to$ VB1 $\to$ VB2 and its conjugate
$\mathcal B_2$: DB1 $\to$ VB4 $\to$ VB5 $\to$ VB6 $\to$ VB4 that give
rise to the circulant matrix structure highlighted in the
checker-board pattern in {\bf (d)} of both imprimitive blocks. {\bf
  (d)} Adjacency matrix of the ideal circuit in {\bf (c)}. We
highlight the two imprimitive blocks ${\mathcal B_1}=$ (VB2, DB3, DB2,
VB1) and ${\mathcal B_2}=$ (DB1, VB4, VB5, VB6) mentioned in the text
and its circulant structure in the normal subgroup ${\bf D_1}$. The
other normal subgroups are also described by circulant blocks and
correspond to imprimitive blocks: ${\mathcal B_3},$ $ {\mathcal B_4},$
$ {\mathcal B_5},$ $ {\mathcal B_6},$ $ {\mathcal B_7}$, as
indicated. Some of these structures also form block-circulant
matrices. Each block of the adjacency matrix $A$
performs a fundamental signal processing task.

\bigskip

FIG. 3. {\bf Symmetry group ${\bf B_{gap}}$ of
  the backward gap-junction circuit.}  {\bf (a)} The real circuits and
{\bf (b)} its adjacency matrix. The symmetry group is factorized as a
direct product of normal subgroups: ${\bf B_{gap}} =\ [{\bf
    C_2}\times {\bf C_2}\times {\bf D_1}]\ \times [{\bf S_{12}}\times
  {\bf D_6} \times {\bf C_2}]$, which leads to a partition of neurons
in two sectors that match the command and motor sectors known
experimentally, as indicated.  {\bf (c)} Ideal circuit and {\bf (d)}
adjacency matrix highlighting the primitive and imprimitive blocks and
their circulant structures from ${\mathcal B_1}$ to ${\mathcal B_9}$.

\bigskip

FIG. 4. {\bf Symmetry groups ${\bf F_{ch}}$
  and ${\bf B_{ch}}$ of the chemical synapse forward and backward
  circuits}.  {\bf (a)} Forward locomotor chemical synapse circuit and
{\bf (b)} its adjacency matrix (ideal circuits, real circuits in 
Supplementary Figs 5 and 6). 
%\ref{fig:forwchemSI} and \ref{fig:backchemSI}).
The symmetry group ${\bf F_{ ch}}$ is factorized into the direct
product of command, motor, and touch subgroups as ${\bf F_{ ch}}
=\ {\bf C_2}\ \times\ [{\bf D_1}]\ \times\ [{\bf S_{10}} \times {\bf
    D_1}]$,
 which, in turn, split up the circuits into independent sectors of
 neurons matching different functions and include also the neuron
 touch class PVC (forward) and AVD (backward). {\bf (c)} The backward
 circuit factorizes as ${\bf B_{ch}}\ =\ {\bf C_2}\ \times\ [{\bf
     C_2}\times {\bf C_2}]\ \times\ [{\bf S_5}\times {\bf S_4}\times
   {\bf S_3}\times {\bf D_1}\times {\bf C_2}\times {\bf C_2}].$ We
 show the ideal circuit and {\bf (d)} its adjacency matrix.
For simplicity we plot only the interneurons that connect to the motor
neurons. Full circuit in SM Fig. 6. 
All neurotransmitters are cholinergic and excitatory (ACh) except for RIM
which uses neurotransmitter Glutamate and Tyramine and AIB which is
glutamatergic (see Supplementary Note 6). 
These different types of synaptic interactions respect the symmetries 
of the circuits, see Supplementary Note 5. 
%SM Section V C. 

FIG. 5. {\bf Symmetry vs other
    methods}. We compare the functional classes obtained from
  symmetries with modularity detection algorithms
  \cite{newman,blondel} and a typical eigenvector centrality
  measure. {\bf (a)} Forward gap-junction circuit classes obtained
  using modularity or community detection detection algorithm from
  \cite{blondel} and {\bf (b)} using eigenvector centrality. {\bf (c)}
  Backward gap-junction circuit modularity and {\bf (d)} eigenvector
  centrality. Both measures, modular detection and centrality, do not
  capture the symmetries and functional classification of this
  connectome.

\clearpage

\begin{figure}[h!]
 \includegraphics[width=.9\textwidth]{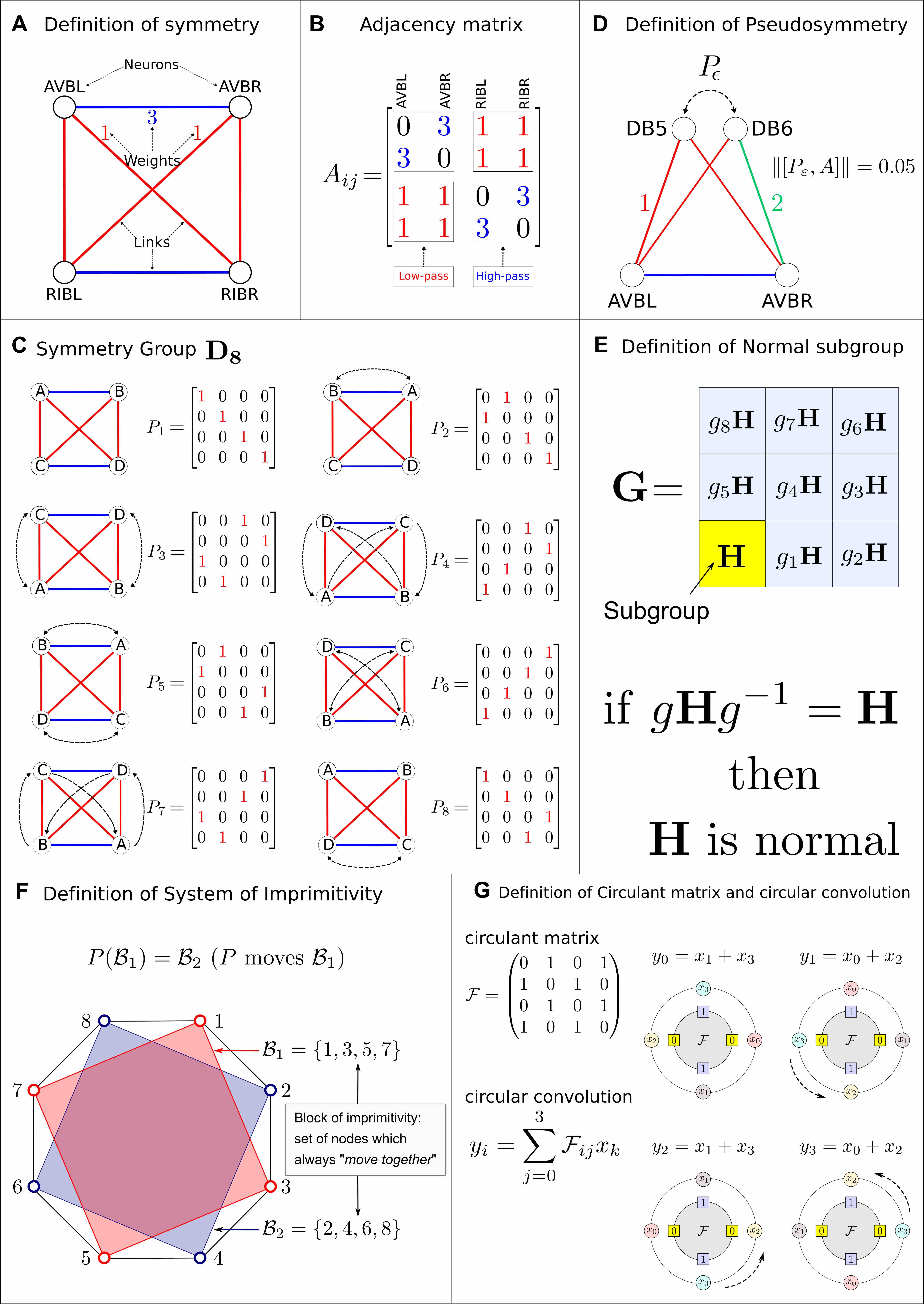} 
 \centering
\caption {}
\label{fig:conceptmap}
\end{figure}

\begin{figure}[h!]
\includegraphics[width=\textwidth]{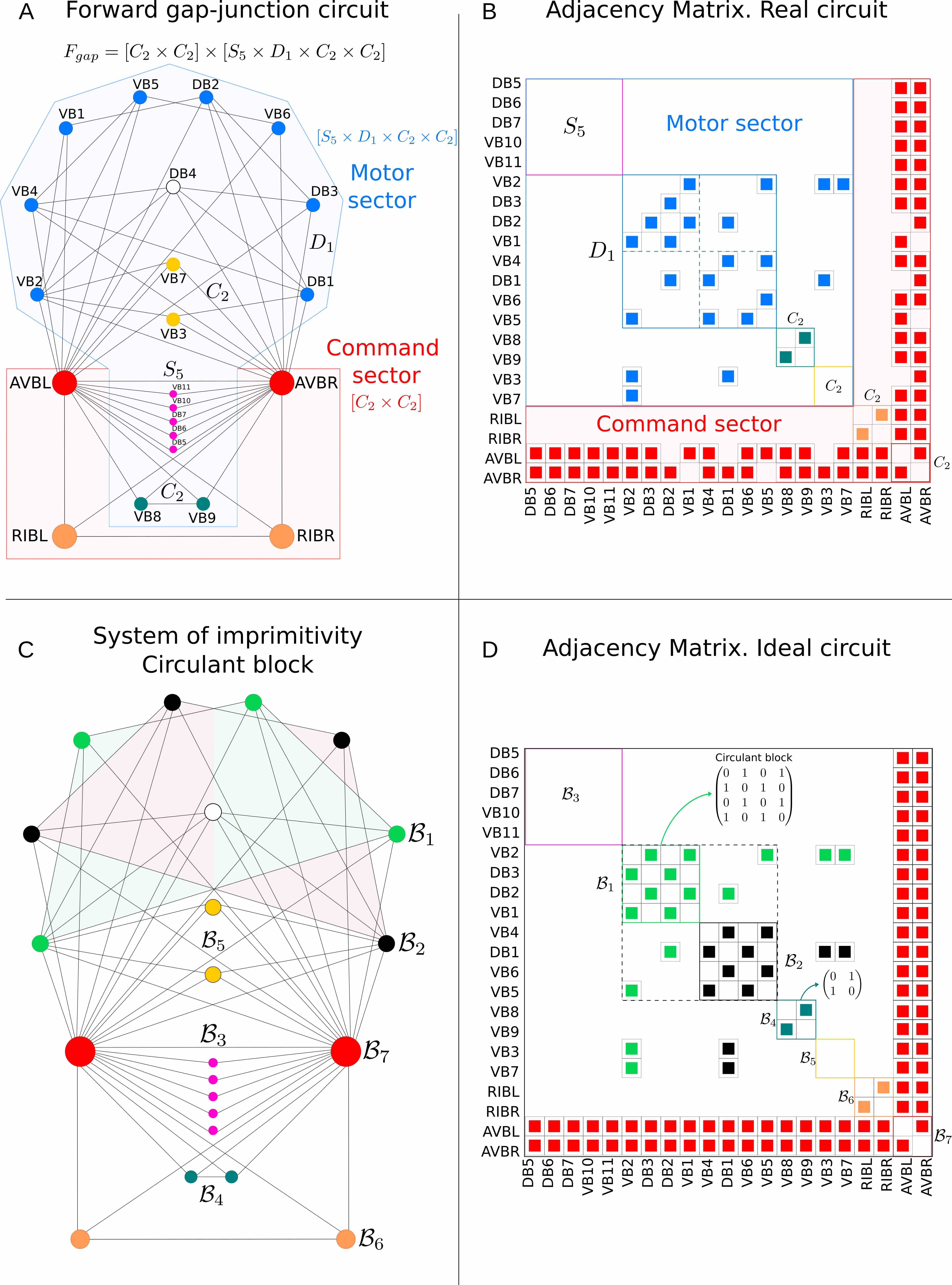} 
\centering
\caption {}
\label{fig:forwardgap}
\end{figure}

\begin{figure}[h!]
\includegraphics[width=\textwidth]{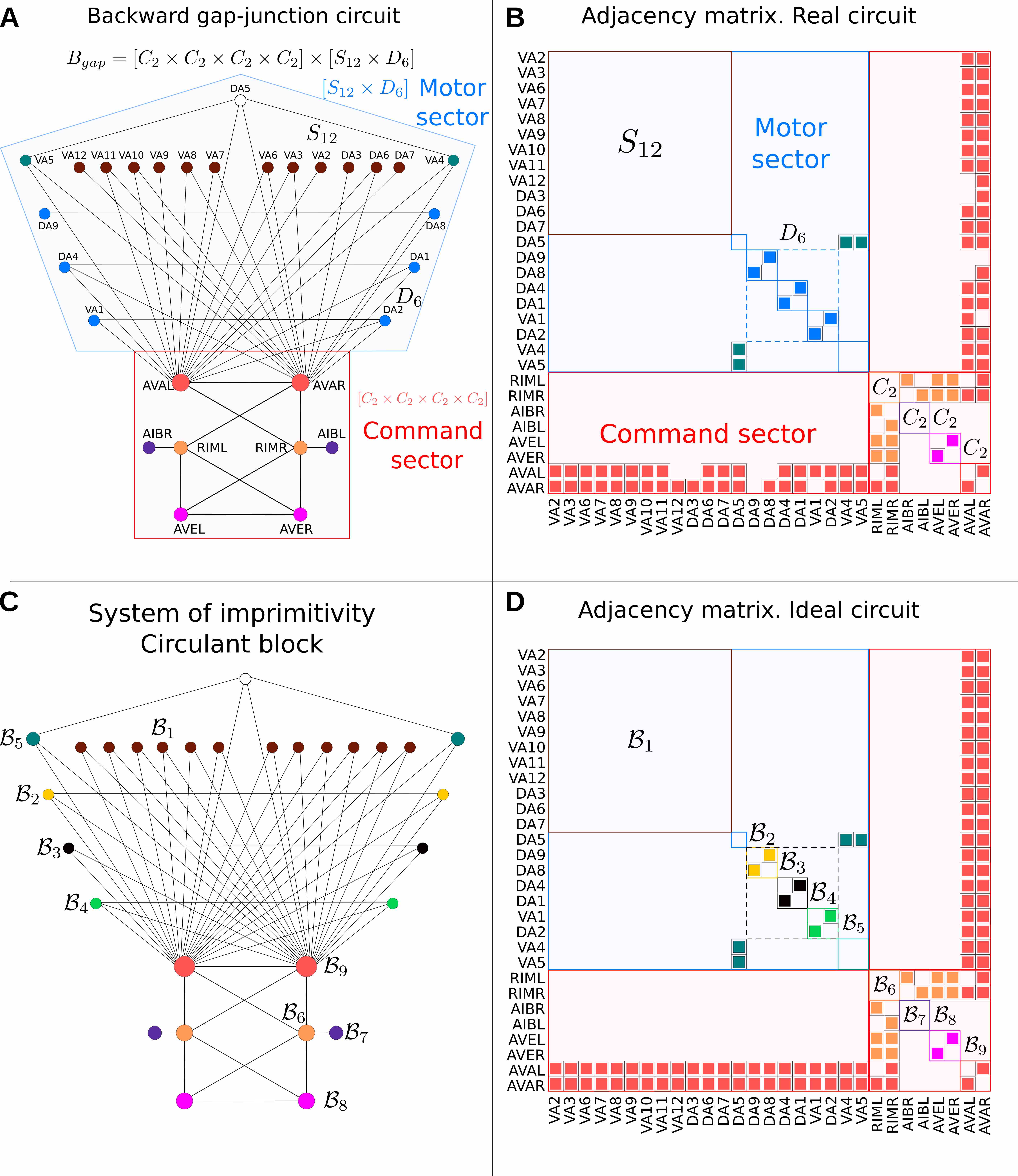} 
\centering
\caption {}
\label{fig:backwardgap}
\end{figure}

\begin{figure}[h!]
\includegraphics[width=\textwidth]{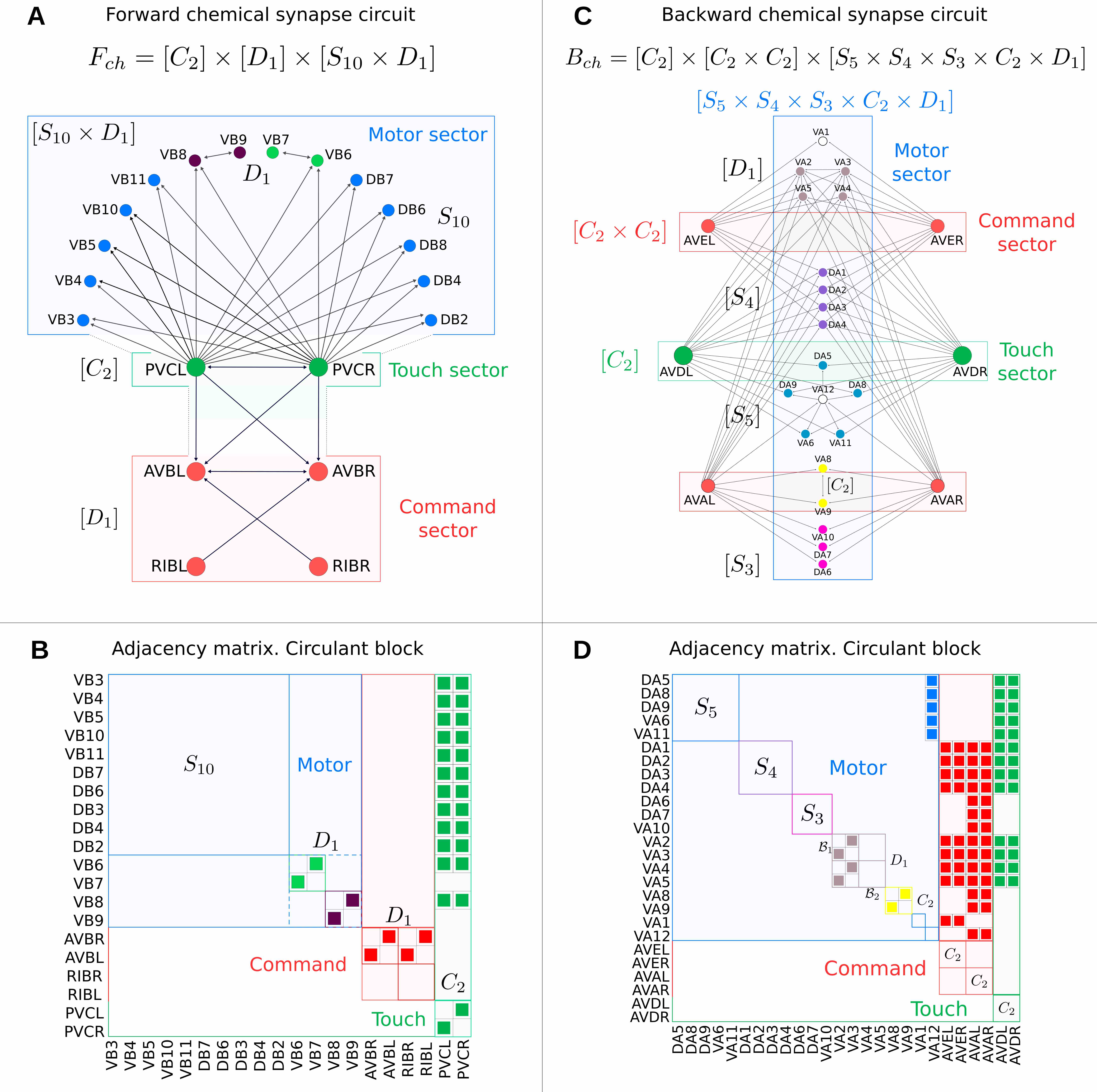} 
\centering
\caption {}
\label{fig:allchemical}
\end{figure}

\begin{figure}[h!]
\vspace{-1.0cm}
 \includegraphics[width=\textwidth]{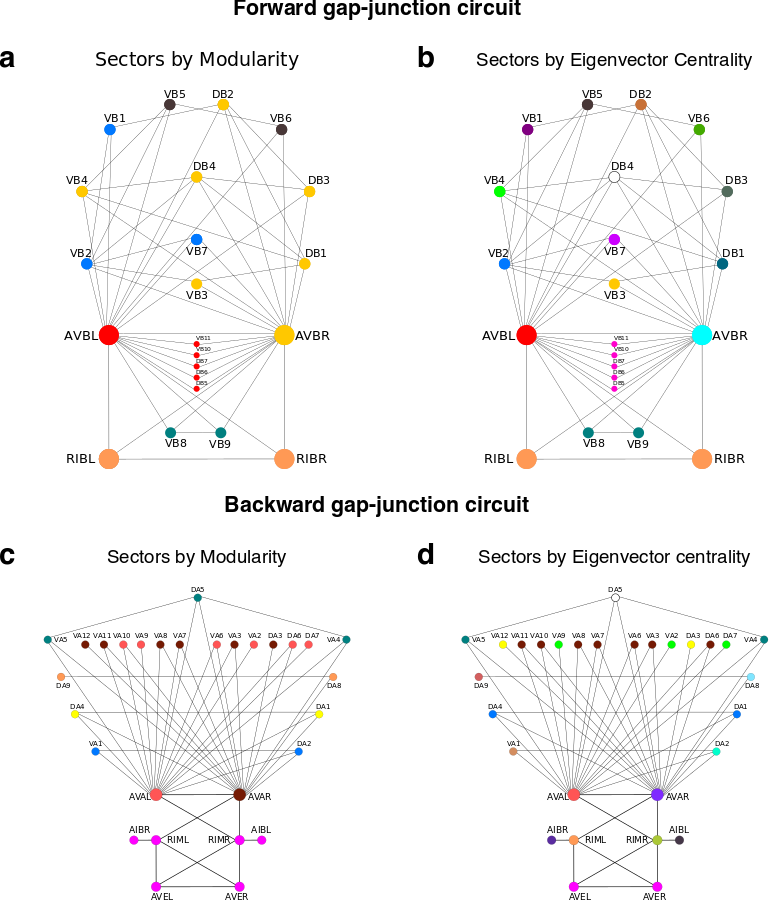} 
\vspace{-0.9cm}
 \caption {}
\label{fig:modularity}
\end{figure}

\begin{table*}[h!]
\centering
\begin{tabular}{ | c | c | c | c |} 
\hline
{\bf Pseudosymmetry - Forward gap-junction} & $\varepsilon$ (\%) & Subgroup &  p-value\\
\hline
\hline
(RIBL, RIBR)  &\ \   0.0\% \ \   &\ \ ${\bf C_2}$ \ \ & 0.001 \\
\hline
(VB3, VB7) &\ \  5.3\% \ \   &\ \ ${\bf C_2}$ \ \  & 0.02 \\
\hline
(VB8, VB9) &\ \  9.6 \% \ \  &\ \ ${\bf C_2}$ \ \  & 0.004\\
\hline
(AVBL, AVBR) &\ \  24.5\% \ \ &\ \ ${\bf C_2}$ \ \ & 0.0007 \\
\hline
(DB5, DB6, DB7, VB10, VB11) &\ \  5.5 \% \ \  &\ \ ${\bf S_5}$ \ \ & 0.0002  \\
\hline
\ \ (DB1, VB2, DB2, VB5, DB3, VB4, VB1, VB6)\ \  &\ \  23.4\% \ \  &\ \ ${\bf D_1}$ \ \ &0.00001  \\
\hline
\hline
{\bf Pseudosymmetry - Backward gap-junction}  & $\varepsilon$ (\%) &\ \ Subgroup\ \ & p-value \\
\hline
\hline
(AIBL, AIBR, RIML, RIMR)    &\ \ 1.5\% \ \  &\ \ ${\bf D_1}$ \ \ & 0.00001 \\
\hline
(DA8, DA9, DA2, VA1, DA1, DA4 ) &\ \ 6.9\% \ \  &\ \ ${\bf D_6}$ \ \ &
$< 10^{-6}$ \\
\hline
(AVEL, AVER) &\ \ 1.5\% \ \  &\ \ ${\bf C_2}$ \ \ & 0.005 \\
\hline
(VA4, VA5)  &\ \ 3.8\% \ \  &\ \ ${\bf C_2}$ \ \ & 0.005 \\
\hline
(VA2, VA3, VA6, VA7, VA8, VA9, VA10, VA11, VA12, 
DA3, DA6, DA7)  &\ \ 13.8\% \ \  &\ \ ${\bf S_{12}}$ \ \ & $<10^{-6}$  \\
\hline 
\hline
{\bf Pseudosymmetry - Forward chemical synapse}  & $\varepsilon$ (\%) & Subgroup & p-value \\
\hline
\hline
(VB3, VB4, VB5, VB10, VB11, DB2, DB4, DB6, DB7, DB8) &\ \ 3.8\% \ \  &\ \ ${\bf S_{10}}$ \ \ & 0.014 \\
\hline
(VB6, VB7, VB8, VB9) &\ \ 3.8\% \ \  &\ \ ${\bf D_1}$ \ \ &0.0012 \\
\hline
(PVCL, PVCR) &\ \ 3.8\% \ \  &\ \ ${\bf C_2}$ \ \ & 0.0006 \\
\hline
(AVBL, AVBR, RIBL, RIBR ) &\ \ 7.6\% \ \  &\ \ ${\bf D_1}$ \ \ & $<10^{-6}$  \\
\hline
\hline
{\bf Pseudosymmetry - Backward chemical synapse}  & $\varepsilon$ (\%) &Subgroup & p-value \\
\hline
\hline
\ \ (VA2, VA3, VA4, VA5)\ \ &\ \ 4.5\% \ \  &\ \ ${\bf D_1}$ \ \ & 0.002 \\
\hline
(VA8, VA9)\ \ &\ \ 0.8\% \ \  &\ \ ${\bf C_2}$ \ \ & $9 \times 10^{-5}$ \\
\hline
(DA5, DA8, DA9, VA6, VA11) &\ \ 10.8\% \ \  &\ \ ${\bf S_5}$ \ \ & $< 10^{-6}$ \\
\hline
(AVAL, AVAR) &\ \ 21.5\% \ \  &\ \ ${\bf C_2}$ \ \ & $4 \times 10^{-6}$ \\
\hline
(AVEL, AVER) &\ \ 15.5\% \ \  &\ \ ${\bf C_2}$ \ \  & $8 \times 10^{-5}$ \\
\hline
(AVDL, AVDR) &\ \ 24.5\% \ \  &\ \ ${\bf C_2}$ \ \ & 0.004 \\
\hline
(DA1, DA2, DA3, DA4)  &\ \ 2.3\% \ \  &\ \ ${\bf S_4}$ \ \ & $4 \times 10^{-6}$  \\
\hline
(VA10, DA6, DA7)\ \  &\ \ 3.8\% \ \  &\ \ ${\bf S_3}$ \ \ & 0.002 \\
\hline
\end{tabular}
\caption{Pseudosymmetries of the locomotion circuit. For each subgroup
  we show the uncertainty constant $\varepsilon$, which is below the
  25\% uncertainty given by the animal to animal experimental
  variability, and therefore the pseudosymmetries have biological
  significance.  The provided p-value indicates that the pseudosymmetries 
  have also statistical significance.}
\label{table:pseudo}
\end{table*}

\clearpage

\renewcommand{\figurename}{ {\bf Supplementary Figure}}
\renewcommand{\tablename}{ {\bf Supplementary Table}}

\setcounter{figure}{0}  

\setcounter{section}{0}

\centerline{ \bf Supplementary Information}

\bigskip

\centerline{\bf Symmetry group factorization reveals the structure-function
  relation in the} \centerline{\bf neural connectome of {\it Caenorhabditis elegans}}

\bigskip

\centerline{ Flaviano Morone and Hern\'an A. Makse}

\bigskip

%\section*{Supplementary Note 1 -
%{\it C. elegans} connectome}
%\label{elegans}

{\bf Supplementary Note 1 - {\it C. elegans} connectome}

\bigskip

We downloaded the most updated connectome of the hermaphrodite worm
{\it Caenorhabditis elegans} ({\it C. elegans}) from the curated
database of Varshney {\it et al.}  \cite{varshney} which is freely
available through the Wormatlas: Altun, Z. F., Hall, D. H. (2002-2006)
Wormatlas \cite{wormatlas}. Available: \url{http://www.wormatlas.org}.
Varshney {\it et al.} report a wiring diagram based on the original
data from White {\it et al.} \cite{white} augmented to include new
serial section electron microscopy reconstructions.  The connectome is
composed of gap junctions which provide direct electrical couplings
between neurons and therefore represent undirected (bidirectional)
links between neurons.  It is also composed of chemical synapses which
use neurotransmitters to transmit signals at the synaptic cleft from a
neuron to a target neuron and are therefore represented by directed
links in the circuits.
Here we consider the circuits of interneurons and motor neurons
involved in two locomotion functions: forward and backward locomotion.
The interneurons connect to motor neurons of classes A and B that
control body wall muscles~\cite{white,chalfie,samuel}.  All neurons
studied here are cholinergic and excitatory (ACh) except for RIM which
uses neurotransmitter Glutamate and Tyramine and AIB which is
glutamatergic (see Supplementary Note 6). 
%SM Section \ref{categories}). 
The different types of synaptic interactions respect the symmetries 
found in the circuits.

\bigskip

{\bf Supplementary Note 2 - Network symmetry group}

\bigskip

A network is a set of nodes $V=\{1,\dots, N\}$ endowed with a
connectivity structure defined by a set of edges $E$ between pair of
nodes.  An edge $i\to j$ is interpreted as an arrow directed from node
$i$ to node $j$, which are said to be connected (or adjacent) to one
another.  The connectivity structure defined by the edge-set $E$ can
be put into the $N\times N$ adjacency matrix $A$, which has nonzero
entries $A_{ij}\neq 0$ only if there is an edge $i\to j\in E$
connecting nodes $i$ to $j$, and $A_{ij}=0$ otherwise. We consider a
weighted adjacency matrix to take into account the number of synaptic
connections as given by \cite{varshney}.

The concept of permutation is as follows.  A permutation of a network,
denoted as $P$, is a bijective map $P:V\to V$ which pairs each node
$i\in V$ with exactly one node $P(i)\in V$, and there are no unpaired
nodes (whence the term bijective map).  As a consequence, any
permutation $P$ has always a well-defined inverse, denoted as
$P^{-1}$. Moreover, since permutations are orthogonal transformation,
we have that $P^{-1}=P^{T}$, where $P^{T}$ denotes the matrix
transpose.  Two permutations $P_1$ and $P_2$ can be composed (or
multiplied), the result being another permutation. Composition of two
permutations is written as $P_1\circ P_2$, and the operation denoted
by $\circ$ is called {\it composition law}. In the following, we omit
for simplicity the symbol $\circ$ and write the composition as
$P_1\circ P_2\equiv P_1P_2$.

A set of permutations ${\bf G}=\{P_1,\dots, P_n\}$ is said to form a 
{\bf permutation group} under composition of its elements if it obeys 
the group axioms~\cite{dixon} listed below. 
{\bf Definition of Permutation Group}:
\begin{enumerate}
\setlength{\itemsep}{0.5pt}
\item existence of the {\bf identity} $I\in {\bf G}$, defined as $I(i)=i$ for all $i$.
\item {\bf associativity} of the composition law : 
$P_i (P_j P_k) = (P_i P_j) P_k$; 
\item {\bf closure} of the composition law: $P_i P_j\in {\bf G}$; 
\item existence of the {\bf inverse} $P_i^{-1}$ for all $P_i\in {\bf G}$, 
defined by $P_i^{-1}P_i=P_iP_i^{-1}=I$.
\end{enumerate}

In a network of size $N$ there are $N!$ different ways to permute its 
nodes. The set of these $N!$ permutations obeys the group axioms listed 
above, so it forms a group. 
However, this is not the symmetry group of the network, because 
not all permutations are, in general, symmetries. 
To qualify as a network symmetry, $P$ must preserve the connectivity
structure, i.e., the network adjacency matrix
$A$~\cite{dixon,pecora1,pecora2,other0}.  In other words, the permuted
adjacency matrix $PAP^{-1}$ must be identical to the original one:
$A=PAP^{-1}$ if $P$ is a permutation symmetry.  Invariance of $A$
under $P$ is formally equivalent to the requirement that $P$ commutes
with $A$, so we have the formal definition of symmetry:
\begin{equation}
[P,A] \equiv PA - AP = 0\ \ \ \iff\ \ \ P {\rm\ is\ a\ permutation\ symmetry}\ .
\label{eq:commutator}
\end{equation}
Permutations which obey Eq.~\eqref{eq:commutator} are formally called 
network automorphisms \cite{dixon}. In short, network symmetry and automorphism 
are synonyms of one another. 
For example, consider the circuit shown in Supplementary 
Fig.~\ref{fig:permutation_elegans}a, and the permutation $P$ acting on 
it represented by the matrix 
\begin{equation}
P =\ \begin{matrix}
\rm AVBL \\
\rm AVBR \\
\rm RIBL \\
\rm RIBR 
\end{matrix}
\left(
\begin{matrix}
    0 &  & 0 & & 1 & & 0 \\
    0 &  & 0 & & 0 & & 1 \\
    1 &  & 0 & & 0 & & 0 \\
    0 &  & 1 & & 0 & & 0 
\end{matrix}
\right)\ ,
\label{eq:legal_perm}
\end{equation} 
which swaps AVBL with RIBL, and AVBR with RIBR. 
This permutation is an automorphism, because the circuits 
before and after the action of $P$ are exactly the same, as seen 
in Supplementary Fig.~\ref{fig:permutation_elegans}a. 
Moreover, it is easy to check that $[P,A]=0$. 
Next, consider the action of the permutation $Q$ shown in 
Supplementary Fig.~\ref{fig:permutation_elegans}b, given by the 
matrix 
\begin{equation}
Q =\ \begin{matrix}
\rm AVBL \\
\rm AVBR \\
\rm RIBL \\
\rm RIBR 
\end{matrix}
\left(
\begin{matrix}
    0 &  & 0 & & 0 & & 1 \\
    0 &  & 1 & & 0 & & 0 \\
    0 &  & 0 & & 1 & & 0 \\
    1 &  & 0 & & 0 & & 0 
\end{matrix}
\right)\ ,
\label{eq:illegal_perm}
\end{equation}
which exchanges AVBL with RIBR and leaves the other neurons fixed. 
Permutation $Q$ is not an automorphism, because it does not 
preserve the connectivity between neurons. Indeed, before the 
action of $Q$, AVBL and AVBR are connected by a link with a weight=3, 
while after they are connected by a link with a weight=1. 
Thus, $Q$ is not a symmetry, because it alters the connectivity 
structure of the circuit by changing the weights on the links. 
Consistently, we also have that $[Q,A]\neq 0$. 

The set of all network automorphisms obeys all group axioms, 
so it forms a group. 
This group, denoted as ${\bf G_{\rm sym}}(A)$, is called the 
{\bf permutation symmetry group of the network}~\cite{dixon}, 
and formally defined as:
\begin{equation}
{\bf G_{\rm sym}}(A) = \{P\ :\ [P,A]=0\}\ .
\label{eq:group}
\end{equation}  
An algorithm to find perfect automorphisms of a given network is call
Nauty, and it is given in Ref.~\cite{nauty}, which is based on the
well-known problem of testing isomorphism of graphs.

\begin{figure}[h!]
 \includegraphics[width=0.7\textwidth]{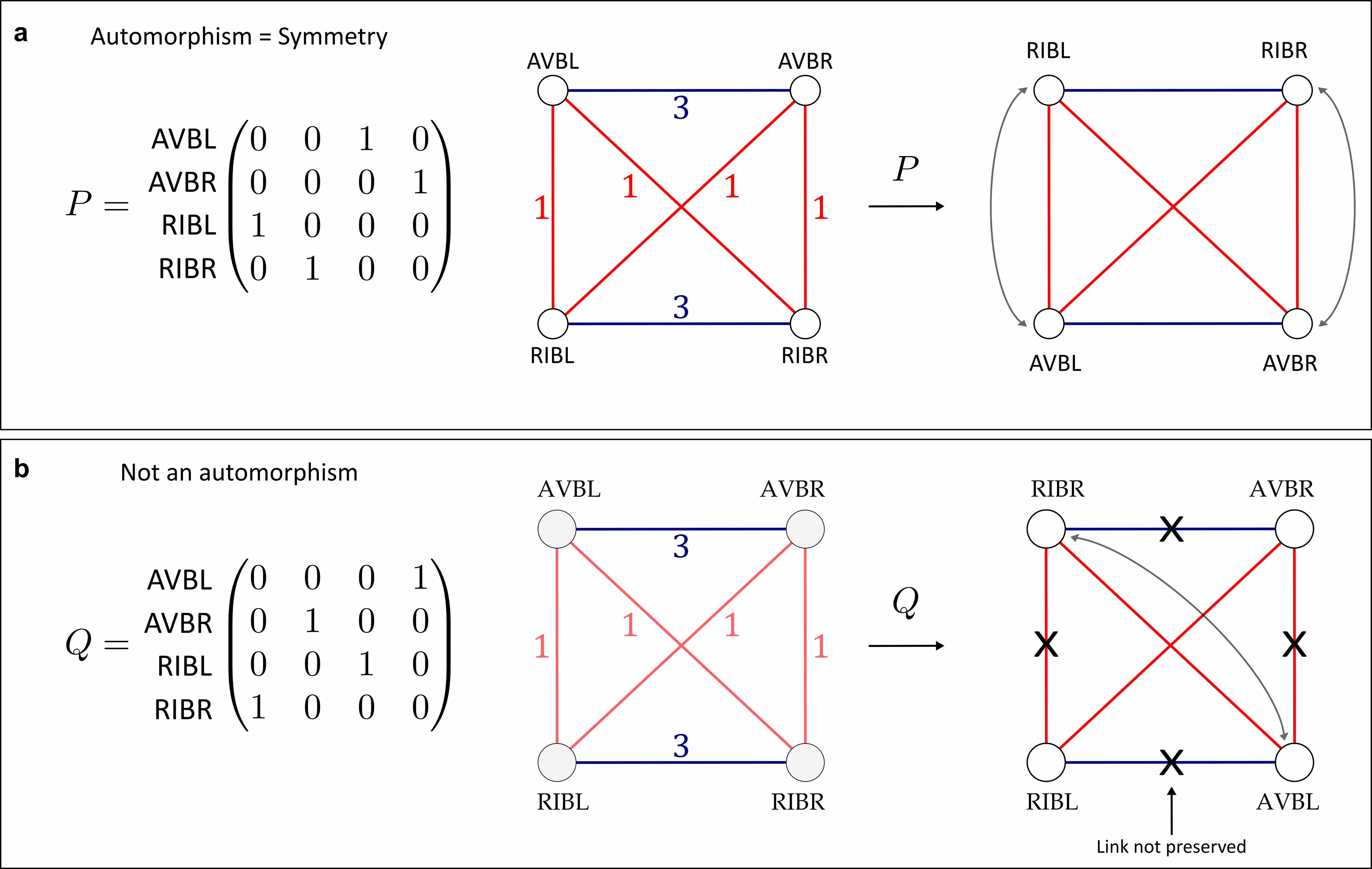} 
\centering
\caption {{\bf Symmetric and non-symmetric permutation}. {\bf (a)}
  Permutation $P$ Eq.~\eqref{eq:legal_perm} is a symmetry of the
  network preserving the connectivity of neurons (automorphism), 
  and commutes with $A$: $[P,A]=0$. 
  {\bf (b)} 
  Permutation $Q$ defined in Eq.~\eqref{eq:illegal_perm} is not a 
  symmetry of the network, because it changes the network connectivity 
  by altering the weights of the links, so it does not commute with $A$: 
  $[Q,A]=0$. }
\label{fig:permutation_elegans}
\end{figure}

\begin{figure}[h]
 \includegraphics[width=0.7\textwidth]{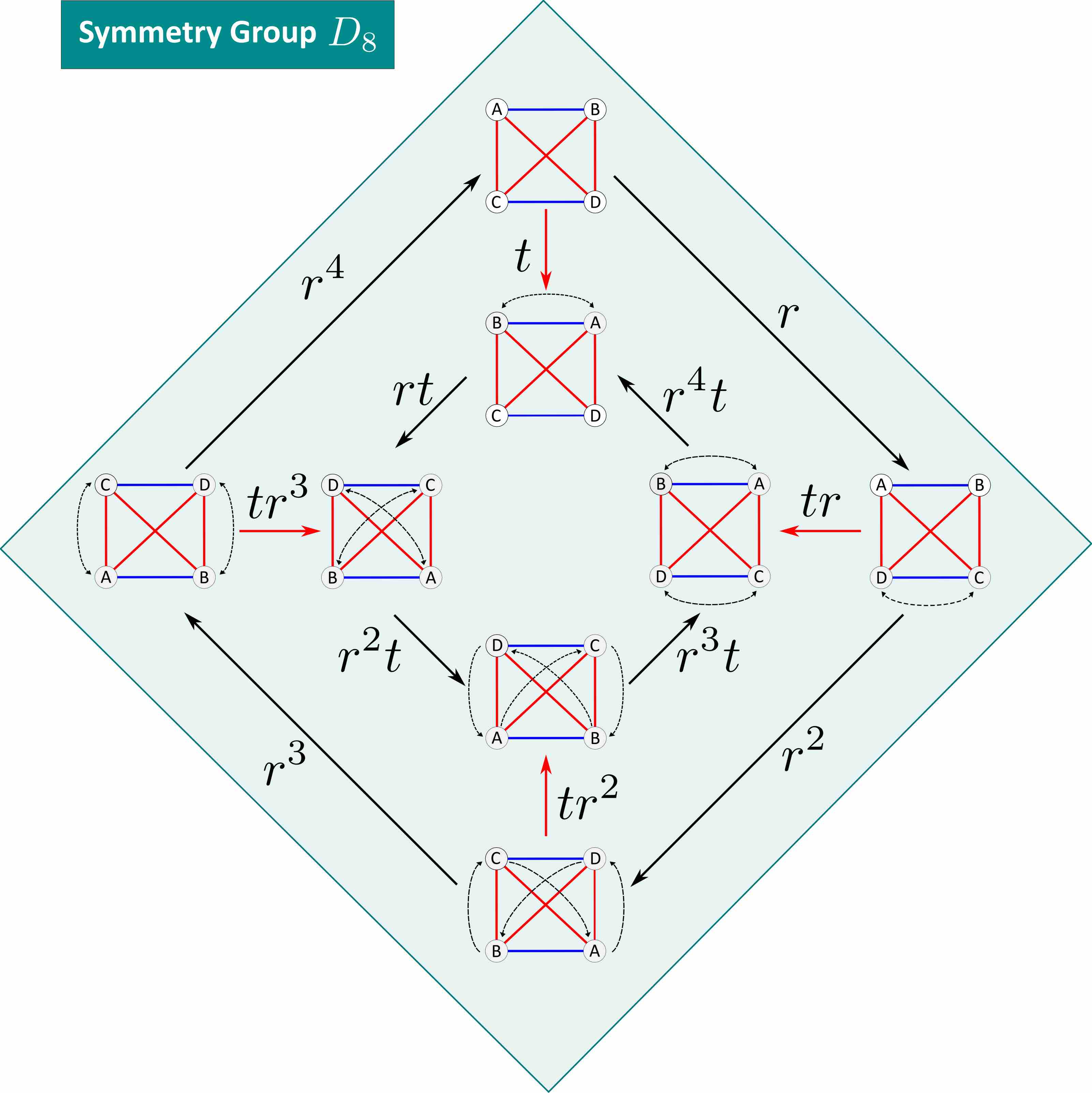} 
\centering
\caption {{\bf Dihedral symmetry group ${\bf D_8}$ of the forward
    gap-junction circuit (interneurons only).} The automorphisms $r$
  and $t$ are the generators of this group, as shown. The structure of
  this group is then converted into the system of imprimitivity when
  this interneuron circuit is incorporated into the whole
  connectome. This is a general property of all functional circuits in
  the connectome, to be elaborated in a follow up paper. }
\label{fig:caleygraph}
\end{figure}

\bigskip

{\bf Supplementary Note 3 - Pseudosymmetries}

\bigskip

%\section{Pseudosymmetries}
%\label{pseudo}

A 25\% variation across animals has been found in the connectivity of
connectomes~\cite{varshney,durbin}.  For this reason, exact symmetries
(= automorphisms) of the connectome are a simplification and an
idealization. However, they should not be regarded as a falsification
of symmetry principles, but rather as an intrinsic property of
biological diversity.  Symmetry principles, in biology, are invariably
idealized and approximate: living systems do have to be sufficiently
non-symmetric to evolve and diversify.  Were it not so, the
nature of exact symmetries would forbid any change in organisms'
structure and functions.  Furthermore, the animal displays a range of
behaviors that are plastic and can change through learning and memory
\cite{rankin}.

Unlike automorphisms, which are canonically defined by
Eq.~\eqref{eq:commutator}, the definition of pseudosymmetry depends
on an additional parameter, a small number $\varepsilon > 0$, which,
for our purposes, represents the 25\% variation existing across
animals.

A permutation $P_{\varepsilon}$ is called a pseudosymmetry if the 
commutator $[P_{\varepsilon},A]$ is non-zero but small
\begin{equation}
||[P_{\varepsilon},A]||\ =\ \varepsilon
%\hat{\varepsilon}\ ,
\label{eq:distanceP}
\end{equation}
that is, $P_{\varepsilon}$ approximates an exact symmetry 
in the limit $\varepsilon\to 0$.

The norm of the commutator in Eq.~\eqref{eq:distanceP}, defined as
\begin{equation}
\Delta(P_{\varepsilon})\ =\ ||[P_{\varepsilon},A]||\ \equiv\ \sum_{i\geq j}|A_{ij}-A_{P(i)P(j)}|\ ,
\label{eq:distance}
\end{equation}
counts the number of links where $P_{\varepsilon}$ and $A$ do not commute. 
The ideal limit of classical symmetry corresponds to
$\Delta(P_{\varepsilon})\to 0$, and we recover exact automorphisms.
In general, the quantity $\Delta(P_{\varepsilon})\to 0$ in
Eq.~\eqref{eq:distance} quantifies the deviation of $P_{\varepsilon}$
from an ideal automorphism. Thus, we are lead naturally to the
following definition of pseudosymmetry.

{\bf Definition of network pseudosymmetry--} A permutation
$P_{\varepsilon}$ is called pseudosymmetry of the network if its
deviation $\Delta(P_{\varepsilon})$ from ideal automorphism is smaller
than a given indetermination constant $\varepsilon$, i.e.,
$\Delta(P_{\varepsilon})< \varepsilon M$, where $M$ is the total
number of links including the weights.  In other words, we require
pseudosymmetries to preserve at least a fraction $(1-\varepsilon)$ of
the total number of links.

\subsection*{Algorithm to find pseudosymmetries}
\label{algorithm}

In the present work, we choose the indetermination constant to be
smaller than $\varepsilon<0.25$, which represents the 25\% variation
in the connectivity of connectomes across
animals~\cite{white,durbin,varshney,hall}, as a condition for
the permutation to be considered a pseudosymmetry. We then obtain the
set of pseudosymmetries shown in the real circuits in the main text.
Finding pseudosymmetries is relatively simple when the size of the
network is small, because they can be determined by an exhaustive
search as those permutations satisfying 
$\Delta(P_{\varepsilon})<M\varepsilon$.
To find the pseudosymmetries we compute for
each permutation $P$ the norm $\Delta(P_{\varepsilon})$ given by
Eq.~\eqref{eq:distance}, and we select only those such that
$\Delta(P_{\varepsilon})<M \varepsilon$. All pseudosymmetries found in
the locomotion circuits represents transformation with indetermination
constant $\varepsilon$ below 25\%. The list of the indetermination
constants of all subgroups appears in Table I.
%~\ref{table:pseudo}.
We notice that pseudosymmetries of locomotion circuits are, in general, 
highly degenerate, and their number increases as a function of $\varepsilon$. 
Due to the fact that $\varepsilon$ is relatively small, these real
circuits can then be easily symmetrized to obtain the circuits with
ideal symmetries with $\varepsilon=0$. This is so, since the
pseudosymmetries are relatively close to a perfectly symmetric
circuit. The provided ideal circuits are examples of idealized
symmetrical circuit and represents the closest ideal structure to the
real one and at the same time respect the same symmetries as the
pseudosymmetries of the real circuit.
The real circuits (and only them) and their pseudo-symmetries remain
the actual circuits to be studied. When the size $N$ of the network is 
larger than $N> 20$, finding pseudosymmetries by using an exhaustive 
search becomes computationally impossible. In this case, pseudosymmetries 
should be determined as the solutions of a constrained quadratic assignment 
problem, to be elaborated and described in detail in a follow up paper.

\bigskip

{\bf Supplementary Note 4 - Factorization of the symmetry group}

\bigskip

%\section{Factorization of the symmetry group}
%\label{sec:factorization}

Factorization of the symmetry group into simple and normal subgroups
is the fundamental tool for understanding the main results of this
work. Descending to subgroups gives us useful information about the
fine structure of the connectome, and eventually will allow us to
identify its basic building blocks.  Next, we explain the notion of
subgroups and then the procedure to find the building blocks of the
connectome through the factorization of its symmetry group.  All
definitions are standard in the group theory literature and appear 
in Ref.~\cite{dixon}.

\medskip

{\bf Definition of Subgroup--} A subset ${\bf H}$ of permutations
selected from a group ${\bf G}$ is said to be a subgroup of ${\bf G}$
if the subset ${\bf H}$ forms itself a group (under the same
composition law that was used in ${\bf G}$). The concept of subgroup
is fundamental in mathematics and physics since it gives the structure
of fundamental forces and particles~\cite{weinberg}.

{\bf Definition of Simple Subgroup--} A simple subgroup is a nontrivial 
group whose only  subgroups are the trivial group and the group itself. 
A group that is not simple can be broken into two smaller groups, a normal 
subgroup and the quotient group, and the process can be repeated, as 
explained next. 

{\bf Definition of Normal Subgroup--} Among all subgroups of a
symmetry group, the normal subgroups, Fig. 1e, are particularly 
significant in this work, since they allow us to define the 
building blocks of the connectome. 
%as shown in the example circuit in Fig. 1f. 
%\ref{fig:conceptmap}f. 
A subgroup ${\bf H}$ is said to be normal in a group ${\bf G}$ if and 
only if ${\bf H}$ commutes with every element $g\in {\bf G}$, i.e., 
$[g, {\bf H}] = g{\bf H} - {\bf H}g=0$ (notice that the requirement is that 
${\bf H}$ commutes with every $g$ as a whole subgroup, not element 
by element).

More precisely, consider a group ${\bf G}$ and a
  subgroup ${\bf H}\leq {\bf G}$.  For a given element $g\in{\bf G}$
  we can form the set $\{gh: h\in{\bf H}\}$, which is called the {\bf
    left coset} of ${\bf H}$ in ${\bf G}$.  Thus we can use ${\bf H}$
  to generate the collection of non-overlapping cosets ${\bf
    H},\ g_1{\bf H},\ g_2{\bf H}, ...$.  Note that while ${\bf H}$ is
  a subgroup, the cosets are, in general, simply sets.  The crux of
  the matter is that if the cosets form themselves a group, then ${\bf
    H}$ is called a normal subgroup. Viceversa, if ${\bf H}$ is a
  normal subgroup, then the cosets do form a group, called the coset
  group.  Next we explain which properties ${\bf H}$ must have in
  order to be a normal subgroup, or equivalently, for the cosets to
  form a group.  Let ${\bf H}$ be a subgroup dividing ${\bf G}$ in
  $N_c$ non-overlapping cosets. Since ${\bf G}$ may be, in general, a
  non-abelian group, the left cosets may differ from right cosets.

  To be definite, in the following we consider only left cosets. Each
  left coset is of the form $g{\bf H}$ for some $g\in{\bf G}$. Let us
  consider two cosets $g_1{\bf H}$ and $g_2{\bf H}$.  Since ${\bf H}$
  is a subgroup, it must contain the identity element $e$,
  i.e. $e\in{\bf H}$. Therefore $g_1e=g_1$ is in the coset $g_1{\bf
    H}$.  Analogously, $g_2e=g_2$ is in the coset $g_2{\bf H}$.  Now,
  if cosets behave like a group, then the product $g_1g_2$ must be in
  the product of two cosets, that is $g_1g_2\in (g_1{\bf H})(g_2{\bf
    H})$.  Since $g_1g_2$ is also in the coset $g_1g_2{\bf H}$, then
  the product of any element in the first coset with any element in
  the second coset should be in the coset $g_1g_2{\bf H}$, i.e.,
  $(g_1{\bf H})(g_2{\bf H})= g_1g_2{\bf H}$.  To see when this
  happens, consider an arbitrary element in the first coset $g_1{\bf
    H}$ and call it $g_1h_1$, and an element in the second coset
  $g_2h_2$. Multiplying these two elements we get $g_1h_1g_2h_2$. If
  this is in the coset $g_1g_2{\bf H}$, then this product must be
  equal to $g_1g_2h_3$ for some $h_3$. Starting from this equation we
  can write:
\begin{equation}
\begin{aligned}
g_1h_1g_2h_2 &= g_1g_2h_3\\
h_1g_2h_2 &= g_2h_3\\
g_2^{-1}h_1g_2h_2 &= h_3\\
g_2^{-1}h_1g_2 &= h_3h_2^{-1}\ .
\end{aligned}
\label{eq:coset}
\end{equation}

Since ${\bf H}$ is a subgroup, the right hand side of Eq.~\eqref{eq:coset} 
is in ${\bf H}$, i.e. $h_3h_2^{-1}\in {\bf H}$. As a consequence, also 
$g_2^{-1}h_1g_2$ is an element of ${\bf H}$, so we have in general 
that $g_2^{-1}{\bf H}g_2 \in {\bf H}$.
In a similar way, one can prove that ${\bf H}\in g_2^{-1}{\bf H}g_2$, 
and thus conclude that 
\begin{equation}
g_2^{-1}{\bf H}g_2 = {\bf H}\ \ \ \to \ \ \ [g_2, {\bf H}]=0\ .
\end{equation}

To recap, we just proved that if ${\bf H}\leq{\bf G}$ is a subgroup 
and the cosets form a group, then it must hold true that $[g, {\bf H}]=0$ 
for any $g\in{\bf G}$. In a similar way it can be proven that the converse 
is also true, that is, if $[g, {\bf H}]=0$ then the cosets form a group. 
If this happens, then ${\bf H}$ is called a {\bf normal} subgroup, denoted 
as ${\bf H}\trianglelefteq {\bf G}$, and the coset group is called 
{\bf quotient} subgroup, denoted as ${\bf G}/{\bf H}$. 
Every group ${\bf G}$ has at least two normal subgroups, which are 
the identity $\{e\}$ and the group itself ${\bf G}$. If these are the only 
normal subgroups then ${\bf G}$ is called a {\bf simple group}. In other 
words, a simple group does not have any quotient subgroups, and for 
this reason simple groups represent the building blocks of other groups. 
Normal subgroups (and only normal subgroups) can be used to 
decompose the symmetry group as a direct product, as we discuss 
next.

{\bf Definition of Direct Product Factorization--} To explain the
factorization of a group as a direct product of normal subgroups, it
is useful to introduce the following notation.  Let us consider a
permutation group ${\bf G}$ and suppose that $K$ is a subset of
$G$. Then, we define the {\bf support} of $K$ by:
\begin{equation}
{\rm supp}(K)=\{ i\in V\ |\ P(i)\neq i\ {\rm for\ at\ least\ one\ } P\in K\}\ .
\end{equation}
Then, suppose that two subsets ${\bf K}$ and ${\bf H}$ of a group
${\bf G}$ have non-overlapping supports, that is ${\rm
  Supp}(K)\cap{\rm Supp}(H)=\emptyset$, then all elements in ${\bf K}$
commute with those in $H$, i.e., $[K,H]=0$.  Assume now that a group
${\bf G}$ can be partitioned into a collection of subsets $\{ {\bf
  H_1, H_2, \times, H_n}\}$ such that for any pair ${\bf H_i}$ and
${\bf H_j}$, $i\neq j$, ${\rm Supp}({\bf H_i})\cap{\rm Supp}({\bf
  H_j})=\emptyset $.  Also, assume that each subset $H_i$ cannot be
further partitioned into smaller subsets with non-intersecting
supports.  The important point is that the subsets ${\bf H_i}$ found
in this way are, by simple construction, the uniquely defined normal
subgroups that factorize ${\bf G}$ into a direct product as:
\begin{equation} 
{\bf G}\ =\ {\bf H_1\times H_2\times \dots H_n}\ .  
\label{eq:directproduct-2}
\end{equation}

More concretely, take the sector of blue motor neurons in Fig. 4a
%Fig.~\ref{fig:allchemical}a 
(VB3, VB4, VB5, VB10, VB11, DB2, DB4, DB6,
DB7, DB8) and its associated subgroup ${\bf S_{10}}$ and the subgroup
${\mathbb T}_{\rm F}^{\rm ch}$ which acts on the sector of touch
neurons colored green PVCL and PVCR.  If we apply any permutation of
${\bf S_{10}}$ to the blue motor neurons, then the neurons PVCL and
PVCR in the other sector are not affected. For instance, a permutation
of VB3 and VB4 is a symmetry that does not affect for instance the
touch sector of interneuron PVCL and PVCR. This factorization is
because VB3 and VB4 are both connected to PVCL and PVCR, and this is a
strong constraint on the connections.  Imagine now that we loss two of
the links and VB3 connects only to PVCL and VB4 only to PVCR.  The
resulting circuit would still be symmetric since we can still permute
VB3 with VB4. But to keep the symmetry of the whole network, this
permutation now triggers the permutation of PVCL and PVCR. Thus, VB3
and VB4 would belong to the touch sector together PVCL and PVCR. We
see how the subgroup structure imposes hard constraints in the network
connectivity.  The fact that the connectivity of the network is
precisely structured to create subgroups which can be factorized is an
interesting result since not all groups possess this
property. Furthermore the factors are aligned with different broad
classification of functions. This is an indication that these
subgroups have biological significance. Thus, the subgroup structure
suggests the segregation of neurons in the network according to
function yet allowing integration since the neurons are connected in
the same circuit.

In Supplementary Note 5
%Section~\ref{group-celegans} 
we will show that both forward and backward circuits, either of gap-junctions 
or chemical synapses, have symmetry groups which factorize as a direct 
product of normal subgroups that correspond to specific broad functional 
categories from the Wormatlas.

\clearpage

{\bf Supplementary Note 5 - Symmetry group of {\it C. elegans} locomotion circuit}

%\bigskip

%\section{Symmetry group of {\it C. elegans} locomotion circuit}
%\label{group-celegans}

\subsection*{ Forward gap-junction circuit}
\label{sec:forwardgap}

The real circuit with the weights of the synapses is shown in
Supplementary Fig.~\ref{fig:forwgapSI}.
The corresponding symmetry group is factorized
as a direct product of 6 normal subgroups:
\begin{equation}
{\bf F}_{\rm gap}\ =\ [{\bf C_2}\times {\bf C_2}]\times
[{\bf S_5}\times {\bf D_1}\times {\bf C_2}\times {\bf C_2}]\ .
\label{eq:Fgap2}
\end{equation}
The pair of subgroups $[{\bf C_2}\times {\bf C_2}]$ acts on the 
set of four interneurons (AVBL, AVBR, RIBL, RIBR), 
but does not move any motor neuron. For this reason, we put 
them together to form the composite subgroup ${\mathbb C}_{\bf F_ { gap}}$, 
which we call {\bf command subgroup of the forward gap-junction circuit} 
and define as:
\begin{equation}
{\mathbb C}_{\bf F_{ gap}}\ = \ {\bf C_2}\times {\bf C_2}\ . 
%{\mathbb C}_{\bf F}^{\rm gap} = {\bf C_2}\times {\bf C_2}
\end{equation}

Similarly, the product $[{\bf S_5}\times {\bf D_1} \times {\bf
    C_2}\times {\bf C_2}]$ in Eq.~\eqref{eq:Fgap2} acts only on the
motor neurons VB and DB, but not on the interneurons.  Therefore, we
put them together to form the composite ${\mathbb M}_{\bf F_{gap}}$, 
and we call it the {\bf motor subgroup of the forward gap-junction circuit}, 
defined as
\begin{equation}
{\mathbb M}_{\rm F_{gap}}\ =\  
[{\bf S_5}\times {\bf D_1} \times {\bf C_2}\times {\bf C_2}]\ .
\end{equation}

The formal decomposition of the circuit into the functional 
categories is:
\begin{equation}
  {\bf F_{ gap}}\ =\ {\mathbb C}_{\bf F_ {gap}}\ \times
  {\mathbb M}_{\bf F_{gap}} \ .
  \end{equation}

\begin{figure}[h!]
 \includegraphics[width=.6\textwidth]{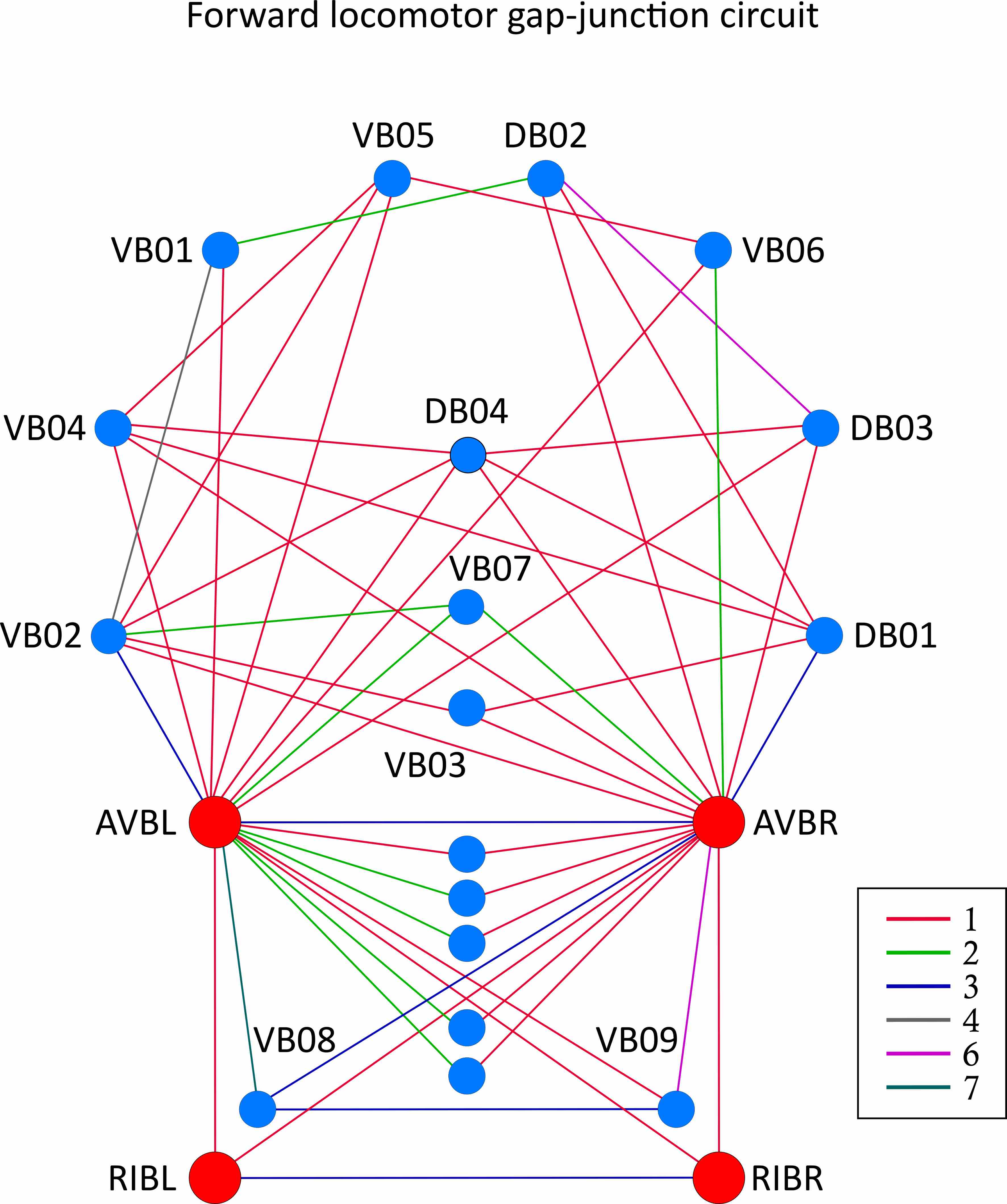}
 \centering
\caption {{\bf Real forward locomotion gap-junction circuit.} 
This circuit comprises 22 neurons divided in 2 sectors: 
the command-sector including the 4 interneurons 
(AVBL, AVBR, RIBL, RIBR); and the motor-sector 
including the remaining motor neurons.}
\label{fig:forwgapSI}
\end{figure}

\subsection*{Backward gap-junction circuit}
\label{backward}
%Similarly to the forward circuit described in
%Section~\ref{sec:forwardgap}, we consider the network made of
%gap-junctions connecting interneurons and motor neurons in the class A
%that drive the backward motion.
The real circuit is shown in Supplementary Fig.~\ref{fig:backgapSI} 
with the weighted links.
The symmetry group of the backward circuit of gap-junctions breaks into a 
direct product of command and motor normal subgroups as: 
\begin{equation}
{\bf B_{gap}}\ =\ ({\bf C_2}\times {\bf C_2}\times {\bf C_2}\times {\bf C_2})
\times ({\bf S_{12}} \times {\bf D_6}\times {\bf C_2} )\ . 
\label{eq:Bgap}
\end{equation}
where the command subgroup is
\begin{equation}
{\mathbb C}_{\bf B_{gap}}\ = \ {\bf C_2}\times {\bf C_2}\times {\bf
  D_1}\ ,
\end{equation}
acts on the command sector (AVAL, AVAR, AVEL,
AVER, RIML, RIMR, AIBL, AIBR), and fix the motor sector, while the
motor subgroup
\begin{equation}
{\mathbb M}_{\bf B_{gap}}\ = \
{\bf S_{12}}\times {\bf D_6}\times {\bf C_2}  ,
\end{equation}
acts only on motor neurons DA and VA and leaves the interneurons
fixed.
The formal decomposition of the circuit is:
\begin{equation}
{\bf B_{gap}}\ =\ {\mathbb C}_{\bf B_{gap}}\ \times
{\mathbb M}_{\bf B_{gap}} \ .
\end{equation}

\begin{figure}[h!]
\includegraphics[width=.6\textwidth]{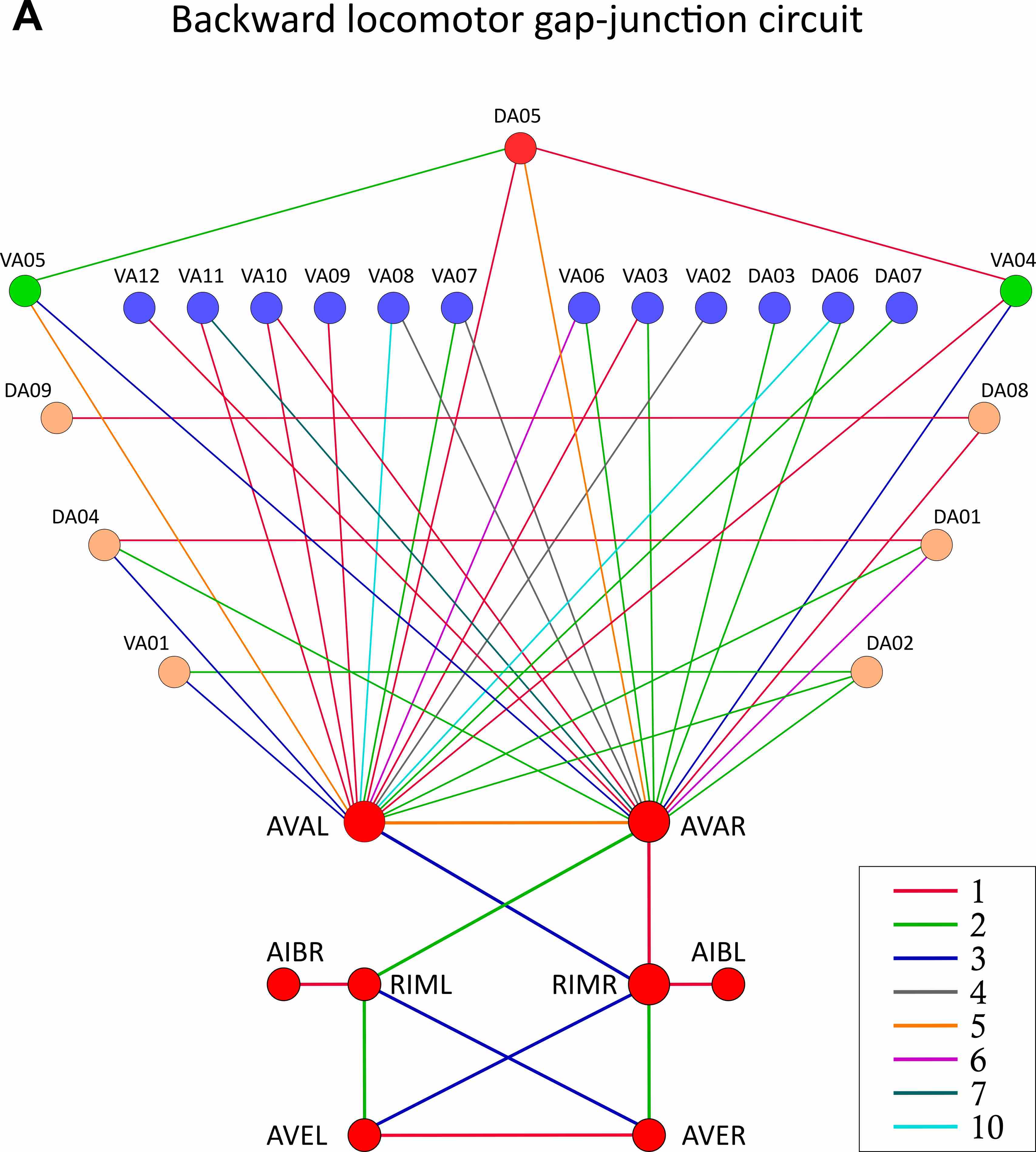} 
\centering
\caption {{\bf Real backward gap-junction circuit.}  
This circuit comprises 29 neurons connected by gap-junctions. 
These neurons form 2 disjoint sectors: the command-sector 
including 8 interneurons (AVAL, AVAR, RIML, RIMR, AIBL, AIBR, 
AVEL, AVER); and the motor sector formed by the remaining 
21 motor neurons. }
\label{fig:backgapSI}
\end{figure}

\subsection*{Forward chemical synapse circuit}
\label{sec:forwardchemical}
We construct the forward circuit of chemical synapses using the same
neurons of the forward gap-junction circuit discussed in 
Supplementary Note 5.
%Section~\ref{sec:forwardgap}. 
In addition, we consider also the two
neurons PVCL and PVCR, since they are connected to the other ones 
via chemical synapses (but not via gap-junctions).  The resulting real
circuit with the weighted links is displayed in Supplementary
Fig.~\ref{fig:forwchemSI}, and its pseudosymmetries are listed in
Table I. 
%~\ref{table:pseudo}.  
We consider the different chemical synaptic
connections according to the different neurotransmitters into
excitatory and inhibitory. All neurons are cholinergic and excitatory
(ACh) except for RIM which uses neurotransmitter Glutamate and
Tyramine and AIB which is glutamatergic, as shown in Supplementary 
Table~\ref{t2}. These different types of synaptic connections do not affect
the symmetries of the circuits and therefore we avoid to plot the type
of neurotransmitter in the links of the chemical synapses circuits for
clarity in all chemical circuits.

The corresponding (pseudo)symmetry group factorizes as the direct
product of five normal subgroups in the following way:
\begin{equation}
  {\bf F_{ ch}}\ =\ ({\bf C_2})\times({\bf D_1}) \times({\bf
    S_{10}}\times {\bf D_1})\ ,
  \label{eq:fchemgroup}
\end{equation}
The first subgroup ${\bf C_2}$ in Eq.~\eqref{eq:fchemgroup} acts 
only on the pair of neurons (PVCL, PVCR) and leaves the rest fixed. 
For this reason, we name it {\bf touch subgroup of forward chemical 
synapse circuit}, nd define as:
\begin{equation}
{\mathbb T}_{\bf F_{ ch}}\ =\  {\bf C_2}\ , \ \ \ {\rm touch\ subgroup.}
\end{equation}
The subgroup ${\bf D_1}$ acts only on the four interneurons, thus
forming a composite subgroup named {\bf command subgroup of the
  forward chemical synapse circuit}, which is defined as:
\begin{equation}
{\mathbb C}_{\bf F_{ ch}}\ =\ {\bf D_1}\ , \ \ \ {\rm
  command\ subgroup.}
\end{equation}
Lastly, the pair of subgroups ${\bf S_{10}}\times {\bf D_1}$ acts 
only on the motor neurons of this circuit, thus forming the 
{\bf motor subgroup of the forward chemical synapse circuit},  
which is defined by: 
\begin{equation}
{\mathbb M}_{\bf F_{ch}}\ =\ [{\bf S_{10}}\times {\bf D_1}]\ ,
\ \ \ {\rm motor\ subgroup.}
\end{equation}

The decomposition of this circuit is:
\begin{equation}
  {\bf F_{ ch}}\ =\ {\mathbb T}_{\bf F_{ch}} \ \times {\mathbb C}_{\bf
    F_{ch}}\ \times {\mathbb M}_{\bf F_{ ch}} \ .
  \label{tcm}
  \end{equation}

\begin{figure}[h!]
 \includegraphics[width=0.6\textwidth]{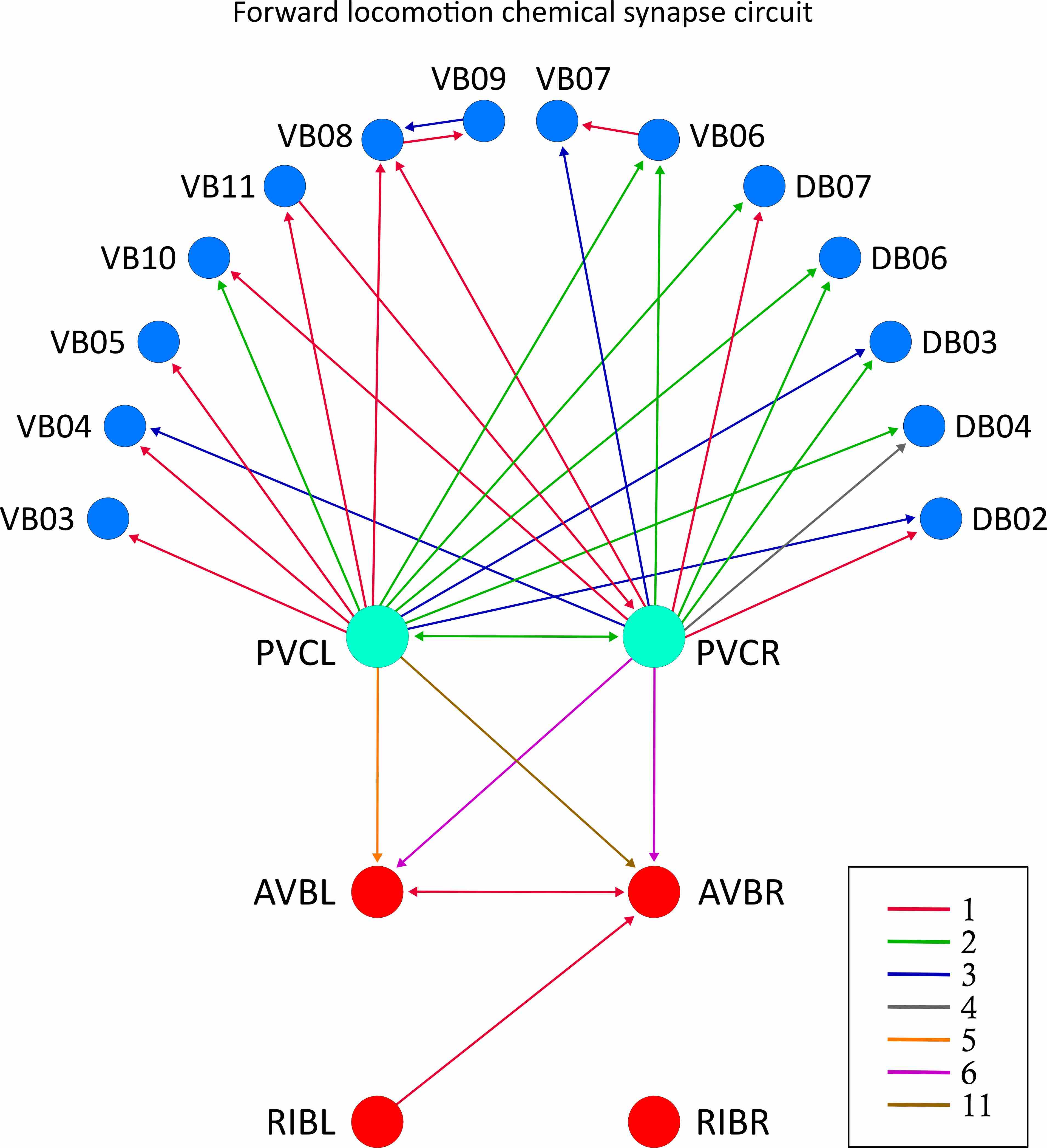} 
\centering
\caption {{\bf Real forward chemical synapse circuit.}  This circuit
  comprises 20 neurons divided in 3 sectors: the touch-sector
  including the pair (PVCL, PVCR); the command-sector including 
  the 4 interneurons (AVBL, AVBR, RIBL, RIBR); and the motor-sector
  including the remaining neurons. All neurons in this circuit are
  cholinergic.}
\label{fig:forwchemSI}
\end{figure}

\begin{figure}[h!]
 \includegraphics[width=\textwidth]{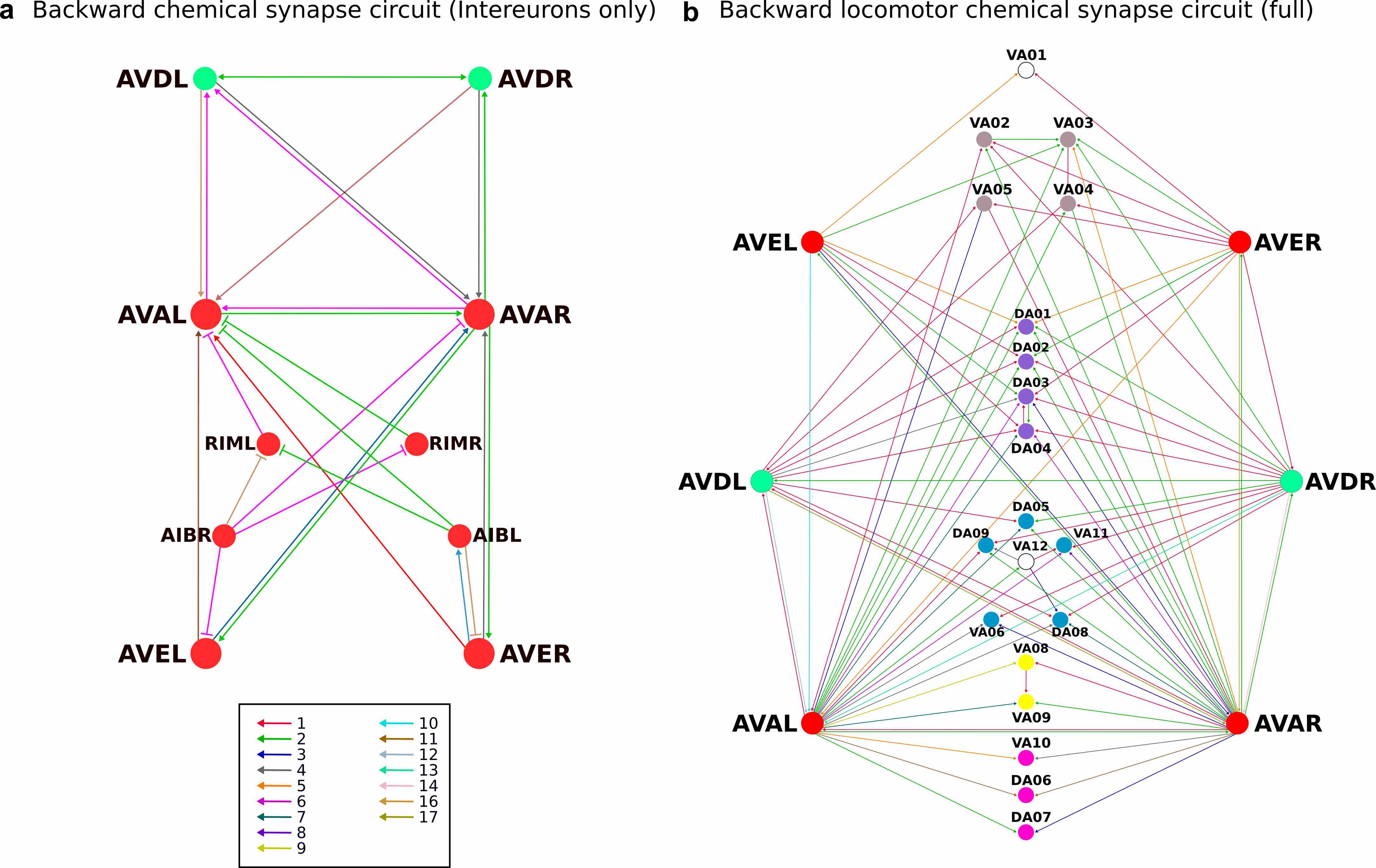} 
\centering
\caption {{\bf Real backward chemical synapse circuit.}  {\bf a}. We
  plot separately the interneurons for clarity. This part of the
  circuit comprises 10 neurons and the chemical synapse between
  them. These neurons form 2 disjoint sectors: the touch-sector
  including the pair (AVDL, AVDR); and the command-sector including
  the other 8 interneurons (AVAL, AVAR, RIML, RIMR, AIBL, AIBR, AVEL,
  AVER). All neurons in this circuit are cholinergic and excitatory (ACh), 
  except for RIM and AIB which are inhibitory: RIM uses neurotransmitter 
  Glutamate and Tyramine and AIB is glutamatergic. The inhibitory nature 
  of their synaptic connections is shown graphically by T-headed arrows 
  ($\dashv$, inhibitory links), as opposed to excitatory synapses represented 
  by ordinary arrows ($\to$, excitatory links). The different types of synapses  
  %neurotransmitters 
  do not affect the pseudosymmetries of this circuit. 
  {\bf b}. We add the motor neurons to the circuit and plot 
  only the interneurons that connect to the motor sector, for clarity. 
  All neurons in this circuit are cholinergic. }
\label{fig:backchemSI}
\end{figure}

For simplicity we plot only the interneurons that connect to the motor
neurons. Full circuit in Supplementary Fig.~\ref{fig:backchemSI}.  
All neurotransmitters are cholinergic and excitatory (ACh) except for RIM
which uses neurotransmitter Glutamate and Tyramine and AIB which is
glutamatergic (see Supplementary Note 6).
%~\ref{categories}). 
These different types of
synaptic interactions respect the symmetries of the circuits, see 
Supplementary Note 5. 

%~\ref{sec:forwardchemical}.

\subsection*{Backward chemical synapse circuit}
\label{sec:backwardchemical}

Since this circuit has a quite dense connectivity structure, for
easier visualization, we plot it by separating two parts. 
Supplementary Fig.~\ref{fig:backchemSI}a shows the real 
circuit involved in the touch-command subgroups.
We then add the motor neurons in the class A and replot the
interneurons involved in backward locomotion but only those that
connect with the motor neurons in Supplementary Fig.~\ref{fig:backchemSI}b. 
These are the neurons AVA, AVE and AVD.  Interneurons AIB and RIM in the
command subgroup are not included for clarity of visualization because
they do not contribute to the connections between the different
sectors.  We then obtain the real circuit displayed in Supplementary 
Fig.~\ref{fig:backchemSI}b involved in the touch-command-motor
subgroups.

The symmetry group of the backward chemical synapse circuit shown 
in Fig. 4c
%~\ref{fig:allchemical}c 
is factorized as:
\begin{equation}
{\rm B}_{\rm ch}\ =\ [{\bf C_2}]\times [{\bf C_2}\times {\bf C_2}]
\times [{\bf S_5} \times {\bf S_4}\times {\bf S_3}\times {\bf
    C_2}\times {\bf D_1}] \ .
\end{equation}
The touch sensitivity subgroup is composed of neurons AVD, the command
interneuron subgroup of neurons AVA, AVE, AIB and RIM, and the motor
subgroup consists of motor neurons VA and DA.  The decomposition of
this circuit is, respectively:
\begin{equation}
  {\bf B}_{\rm ch}\ =\ {\mathbb T}_{\bf B_ {ch}}\ \times
  {\mathbb C}_{\bf B_{ch}} \times
  {\mathbb M}_{\bf B_{ch}} \ .
  \end{equation}

\clearpage

{\bf Supplementary Note 6 - Wormatlas functional categories on neurons}

\bigskip

%\section{Wormatlas functional categories on neurons}
%\label{categories}

Broad functional categories of neurons are provided at the Wormatlas:
\url{http://www.wormatlas.org/hermaphrodite/nervous/Neuroframeset.html},
Chapter 2.2 \cite{wormatlas}.  A classification for every neuron into
four broad neuron categories is provided as follows: (1) {\it 'motor
  neurons, which make synaptic contacts onto muscle cells'}, (2) {\it
  'sensory neurons',} (3) {\it 'interneurons, which receive incoming
  synapses from and send outgoing synapses onto other neurons'}, and
(4) {\it polymodal neurons, which perform more than one of these
  functional modalities'}.

The Wormatlas classifies most neurons (some of them unknown) in
further functional categories as well as provides the
neurotransmitters. We reproduce the information from the Wormatlas
used in the main text in Supplementary Table~\ref{t1} and 
Supplementary Table~\ref{t2}.

\begin{table}[ht!]
\centering
\begin{tabular}{ | c | c | c | c |}
  \hline
  \multicolumn{4}{|c|}{Forward circuit}\\
    \hline
  \hline Neuron & Functional Category & Explanation & Neurotransmitter\\
  \hline \hline
  AVB & interneuron & driver cell for forward locomotion & ACh
  \\
  \hline
  RIB & interneuron/motor & second layer interneuron, & ACh \\
      & polymodal            & process of integration of
  information, locomotion & \\
  \hline
  PVC & interneuron & command interneuron for forward locomotion,  & ACh \\
   &  & 
  modulates response to harsh touch to tail &  \\
  \hline
  VB & motor (sensory) & locomotion (ventral), proprioception & ACh \\
  \hline
  DB & motor  & forward locomotion (dorsal), proprioception & ACh \\
  \hline
  \hline
\end{tabular}
\caption{Functional categories of the neurons in the forward circuit according to the Wormatlas.}
\label{t1}
\end{table}

\begin{table}[ht!]
\centering
\begin{tabular}{ | c | c | c | c |}
  \hline
  \multicolumn{4}{|c|}{Backward circuit}\\
  \hline     \hline Neuron & Function category & Explanation & Neurotransmitter\\
  \hline \hline
  AVA & interneuron & command interneuron, locomotion,
   & ACh
  \\
     & & 
  driver cell for backward locomotion & 
  \\
  \hline
    AVE & interneuron & command interneuron,  & ACh
  \\
     &  &
  drive backward movement & 
  \\
  \hline
  RIM & interneuron & second layer interneuron,  & Glu, Tyr \\
   & (motor) & process of integration of
  information, locomotion&  \\
  \hline
  &   & first layer amphid interneuron, &  \\
 AIB  & interneuron  & locomotion, food and odor-evoked behavior, & Glu \\
 &  & lifespan, starvation response &  \\
 \hline
  AVD & interneuron & command interneuron,  & ACh \\
    &  & modulator for backward locomotion induced by head-touch & \\
  \hline
  VA & motor & locomotion & ACh \\
  \hline
  DA & motor  & backward locomotion  & ACh \\
  \hline
  \hline
\end{tabular}
\caption{Functional categories of the neurons in the backward circuit according to the Wormatlas.}
\label{t2}
\end{table}

\bigskip

{\bf Supplementary Note 7 - Blocks of imprimitivity}

\bigskip

%\section{Blocks of imprimitivity}
%\label{sec:imprimitivity}

The correspondence of network building blocks and simple subgroups
provides a rigorous theoretical characterization of the network
connectivity structure and a natural interpretation of its broad
functional categories according to the Wormatlas.  However, a more
accurate description of functionality should take into account also
the splitting of these building blocks into finer topological
structures.  The fine structure corrections to the building blocks can
be obtained systematically through the concept of {\bf system of
  imprimitivity} of a symmetry group ${\bf G}$. All definitions appear
in~\cite{dixon}.

To define a system of imprimitivity we need first the notions of 
{\bf transitivity} and {\bf blocks}. 
A group ${\bf G}$ is said to be {\bf transitive} on the set of nodes $V$ if 
for every pair of nodes $i, j\in V$ there exist $P\in {\bf G}$ such that 
$P(i)=j$ (in other words, ${\bf G}$ has only one orbit). 
A group which is not transitive is called intransitive. 
A block is defined as a non-empty subset ${\mathcal B}$ of nodes 
such that for all permutations $P\in {\bf G}$ we have that: 
\begin{itemize}
\item either $P$ fixes ${\mathcal B}$: 
$P({\mathcal B}) = {\mathcal B}$; 
\item or $P$ moves ${\mathcal B}$ completely: 
$P({\mathcal B})\cap {\mathcal B} = \emptyset$. 
\end{itemize}
If ${\mathcal B} = \{i\}$ or ${\mathcal B} = \{V\}$, then ${\mathcal
  B}$ is called a trivial block. Any other block is nontrivial. If
{\bf G} has a nontrivial block then it is called {\bf imprimitive},
otherwise is called {\bf primitive}.

The importance of blocks rests on the following fact.  
If ${\mathcal B}$ is a block for ${\bf G}$, then $P({\mathcal B})$ is 
also a block for every $P\in {\bf G}$, and is called a conjugate block 
of ${\mathcal B}$. Suppose that ${\bf G}$ is transitive on the set of
nodes $V$ and define $\Sigma = \{ P({\mathcal B})\ |\ P\in {\bf G}\}$
as set of all blocks conjugate to ${\mathcal B}$. Then the sets in
$\Sigma$ form a partition of the set of nodes $V$, and each element of
$\Sigma$ is a block for ${\bf G}$. We call $\Sigma$ a {\bf system of
  imprimitivity} for the (symmetry) group ${\bf G}$ \cite{dixon}.

In the text we have shown that  the action of ${\bf G}$ on the system of
imprimitivity $\Sigma$ gives important information about the
functionality of the neural circuits, provided ${\mathcal B}$ is not a
trivial block.

\bigskip

{\bf Supplementary Note 8 - Circulant Matrices and Fast Fourier Transform}

\bigskip

%\section{Circulant Matrices and Fast Fourier Transform}
%\label{kernel}
%
In this section we discuss the relationship between circulant 
matrices and discrete Fourier analysis (see Fig. 1g).
%~\ref{fig:conceptmap}g). 
In particular, we show that the eigenvalues of circulant matrices can 
be computed extremely fast through a routine of just $O(N\log N)$ 
operations, called Fast Fourier Transform (FFT). 

We start the discussion by recalling that a  circulant matrix 
$A={\rm circ}(a_0,a_1,\dots, a_{N-1})$ can always be written as 
a polynomial of the permutation matrix $P={\rm circ}(0,1,0,\dots, 0)$
of degree at most $N-1$, that is: 
\begin{equation}
A = a_0I + a_1P + a_2P^2 + ... + a_{N-1}P^{N-1}\ . 
\end{equation}
For instance, the low-pass filter: 
\begin{equation}
  {\mathcal L} = {\rm circ}(1, 1) =
\begin{bmatrix} 1& &1\\ 1&&1
\end{bmatrix}, 
\end{equation}
can be written as ${\mathcal L} = I + P$. 
%Eq. (\ref{lowpass}) can be written 
%as $A = P+P^2$.
Next, we introduce the matrix $F$ with entries $F_{ab}$ defined 
as follows:
\begin{equation}
F_{ab} = \frac{1}{\sqrt{N}}\ e^{\frac{2\pi i}{N}ab}\ .
\label{eq:fft}
\end{equation}
Matrix $F$ is a unitary matrix ($ F^{\dagger}=F^{-1}$) which 
represents the kernel of the discrete Fourier transform (DFT). 
Specifically, given a vector $x$, its DFT, denoted as $\tilde{x}$, 
is the vector defined as: $\tilde{x}_a=\sum_bF_{ab}v_b$. 
The crucial point is that the permutation matrix $P={\rm circ}(0,1,0,\dots, 0)$ 
is diagonalized by $F$, that is $P = F\Lambda F^{-1}$. This 
can be easily seen by calculating explicitly the product $F^{-1}PF$, 
which reads:
\begin{equation}
\begin{aligned}
(F^{-1}PF)_{ab} = 
\frac{1}{N}\sum_{k=0}^{N-1}\sum_{m=0}^{N-1} 
e^{-\frac{2\pi i}{N}ak}P_{km}e^{\frac{2\pi i}{N}mb} = 
\frac{e^{\frac{2\pi i}{N}b}}{N}\sum_{k=0}^{N-1} e^{\frac{2\pi i}{N}k(b-a)} = 
\delta_{ab}\ e^{\frac{2\pi i}{N}b}\ .
\end{aligned} 
\label{eq:fftP}
\end{equation}
As a consequence of Eq.~\eqref{eq:fftP}, any circulant matrix 
$A$ is also diagonalized by $F$ as
\begin{equation}
(F^{-1}AF)_{ab} =  \sum_{n=0}^{N-1}a_n(F^{-1}P^nF)_{ab} = 
\delta_{ab}\sum_{n=0}^{N-1}a_n e^{\frac{2\pi i}{N}nb}\ ,
\end{equation}
so we can write down the eigenvalues $\{\lambda_a\}$ of $A$ as
\begin{equation}
\lambda_a = \sum_{n=0}^{N-1}a_n e^{\frac{2\pi i}{N}na}\ , \ \ \ a = 0,\dots,N-1\ .
\end{equation}
Eigenvalues $\{\lambda_a\}$ can be computed efficiently using 
the FFT of the vector 
$\vec{\alpha}\equiv \frac{1}{\sqrt{N}}(a_0, a_{N-1}, ..., a_1)^T$. 
To see this, we rewrite $\lambda_a$ as 
\begin{equation}
\begin{aligned} 
\lambda_a &= \sum_b (F^{-1}AF)_{ab} = \sum_{bk}(F^{-1}A)_{ak}F_{kb} =
\sqrt{N}\sum_{k}(F^{-1}A)_{ak}\delta_{k0} \\
&= \sqrt{N}(F^{-1}A)_{a0}= \frac{1}{\sqrt{N}} \sum_b F_{ab}A_{b0}\ , 
\end{aligned} 
\label{eq:ffteasy}
\end{equation}
where we used the fact that $F$ satisfies the following sum rules: 
\begin{equation}
\begin{aligned} 
\sum_{b=0}^{N-1} F_{ab} &= \sqrt{N}\delta_{a0}\ ,\\
\sum_{b=0}^{N-1} F^{-1}_{ab} F_{b0} &=  \frac{1}{N} \delta_{a0}\ .
\end{aligned} 
\end{equation}
Using the vectors $\vec{\alpha}\equiv \frac{1}{\sqrt{N}}(a_0, a_{N-1}, a_{N-2}, ..., a_1)^T$ 
and $\vec{\lambda}\equiv (\lambda_0, \lambda_1, ..., \lambda_{N-1})^T$,  
we can write Eq~\eqref{eq:ffteasy} in the simple form
\begin{equation}
F\vec{\alpha} =  \vec{\lambda}\ ,
\end{equation}
which shows that the eigenvalues $\{\lambda_a\}$ of $A$ are the
components of the DTF of vector $\vec{\alpha}$.  Since $F\vec{\alpha}$
can be evaluated in $O(N\log N)$ operations using a FFT, then the
computational effort for diagonalizing a circulant matrix $A$ requires
$O(N\log N)$ operations, too. Thus, we can interpret the functionality
of the circulant matrix as a fast way (almost linear in the number of
nodes) to perform a Fourier Transform for processing of information.

\end{document}